\documentclass[ twocolumn, pra, superscriptaddress, amsmath, showpacs,
tightenlines,twoside,twocolumn]{revtex4}%
\usepackage{amsfonts}
\usepackage{graphicx}
\usepackage{amstext}
\usepackage{amsmath}
\usepackage{amssymb}
\usepackage{latexsym}
\usepackage{hyperref}
\usepackage{times}
\usepackage{subfigure}
\usepackage{color}%
\providecommand{\U}[1]{\protect\rule{.1in}{.1in}}
\begin{document}
\title{Engineering entangled microwave photon states through multiphoton interactions
between two cavity fields and a superconducting qubit}
\author{Yan-Jun Zhao}
\affiliation{Institute of Microelectronics, Tsinghua University, Beijing, 100084, China}
\author{Chang-Qing Wang}
\affiliation{Institute of Microelectronics, Tsinghua University, Beijing, 100084, China}
\author{Xiaobo Zhu}
\affiliation{Institute of Physics, Chinese Academy of Sciences, Beijing, 100190, China}
\author{Yu-xi Liu}
\email{yuxiliu@mail.tsinghua.edu.cn}
\affiliation{Institute of Microelectronics, Tsinghua University, Beijing, 100084, China}
\affiliation{Tsinghua National Laboratory for Information Science and Technology (TNList),
Beijing 100084, China}
\keywords{entanglement, sideband transition, superconducting qubit}
\pacs{42.50.Dv, 42.50.Pq, 74.50.+r}

\begin{abstract}
It has been shown that there are not only transverse but also longitudinal
couplings between microwave fields and a superconducting qubit with broken
inversion symmetry of the potential energy. Using multiphoton processes
induced by longitudinal coupling fields and frequency matching conditions, we
design a universal algorithm to produce arbitrary superpositions of two-mode
photon states of microwave fields in two separated transmission line
resonators, which are coupled to a superconducting qubit. Based on our
algorithm, we analyze the generation of evenly-populated states and NOON
states. Compared to other proposals with only single-photon process, we
provide an efficient way to produce entangled microwave states when the
interactions between superconducting qubits and microwave fields are in the
ultrastrong regime.

\end{abstract}
\revised{\today}

\startpage{1}
\endpage{ }
\maketitle

\section{Introduction}

Superconducting transmission line resonators can be used as quantum data
buses, quantum memories, and single microwave photon
detectors~\cite{Wallraff2004,Blais2004}. They usually work\ in the microwave
regime and can also be used as quantum nodes in so-called quantum
networks~\cite{Kimble2008,Ritter2012}. It is well known that the entanglement
is one of the most important resources for quantum information
processing~\cite{Nielsen2010}, and microwave photons play a critical role in
quantum state control for solid state quantum devices. Therefore, engineering
arbitrarily entangled microwave photon states is a very fundamental issue for
both solid state quantum information processing and quantum
optics~\cite{Scully1997} on superconducting quantum chips.

Usually, nonclassical photon states of a single-mode cavity field are
generated through the interaction between the cavity field and the two-level
atom. The methods of generating nonclassical photon states can be classified
into two ways. One is to engineer appropriate Hamiltonians in different
evolution durations by tuning experimental parameters when the target state is
being generated~\cite{YXLiu-EPL,YXLiu-PRA,Martinis-1,Martinis-2}. The other
one is to obtain the target state via appropriately designed
measurements~\cite{Vogel1993}. The former one is deterministic, while the
latter one is probabilistic and usually has a low probability to succeed. If
the nonclassical state is generated using natural atomic systems, the latter
method is usually more practical since most of parameters are not possible or
not easy to be tuned. However, in artificial atomic systems, the former method
is more appropriate because system parameters can be artificially controlled.
For example, superconducting quantum circuits
(SQCs)~\cite{r1,r2,r3,r4,r5,r6,r7,r8} provide us a very convenient way to
deterministically engineer nonclassical states of a single-mode microwave
field by varying the system
parameters~\cite{YXLiu-EPL,YXLiu-PRA,Martinis-1,Martinis-2}.

The method of deterministically generating entangled photon states using
atomic systems can be tracked to that of generating entangled phonon states of
two vibrational modes~\cite{Gardiner1997}, in a trapped ion interacting with
laser fields, by using different sideband transitions. However, the number of
steps in such a method~\cite{Gardiner1997} exponentially depends on the
maximum phonon numbers. A few proposals were put forward to overcome the
exponential dependence of the phonon number by introducing auxiliary atomic
energy levels~\cite{Drobny1998,Zheng2000}, using phonon number dependent
interactions~\cite{Kneer1998}, or employing multiphonon transitions of high
phonon numbers~\cite{Zou2002,Zheng2000}. These methods have successfully
reduced the number of steps into quadratic polynomials of the maximum phonon numbers.

The generation of entangled microwave photon states of two modes using
superconducting qubit has been
studied~\cite{Strauch2010,Strauch2012PRA,Sharma2015}, where a classically
driven superconducting qubit with time-dependent frequency is coupled to two
microwave fields in two separated cavities. The interaction Hamiltonian
between the superconducting qubit and the cavity fields of two modes is
described by the Jaynes-Cummings model. Therefore, there is only single photon
transition in each step. However, the photon-number-dependent Stark
effects~\cite{Strauch2010,Strauch2012PRA,Sharma2015} induced by the
qubit-field coupling make it possible to independently implement operations
for photon states. Thus, the number of steps also quadratically depends on the
maximum photon number.

It has been shown that the superconducting qubit and the cavity field can have
both transverse and longitudinal couplings when the inversion symmetry of the
qubit potential energy is broken~\cite{Liu-PRL,Liu-NJP}. The longitudinal
coupling can induce multiphoton transitions~\cite{Zhao2015} in different
sidebands as in trapped ions~\cite{trappedions,wei} and thus arbitrary photon
states of a single-mode cavity field can be more conveniently
engineered~\cite{Zhao2015}. Motivated by
studies~\cite{Strauch2010,Strauch2012PRA,Sharma2015,Liu-PRL,Liu-NJP,Zhao2015},
we study a method to generate entangled microwave photon states in two
separated cavities coupled by a superconducting qubit using multiphoton
transitions. We first show that the longitudinal couplings can induce two-mode
multiphoton processes similar to those in trapped ions~\cite{Steinbach1997},
and then study an efficient way to generate superposed two-mode photon states.

The paper is organized as below. In Sec.~\ref{sec:inter-mode}, an effective
Hamiltonian, similar to that of trapped ions with two vibrational
modes~\cite{Steinbach1997}, is derived, and then different sideband
transitions are discussed. In Sec.~\ref{sec:algorithm}, a new algorithm is
introduced to generate arbitrary superpositions of two-mode photon states. In
Sec.~\ref{sec:ParSel}, we discuss how to choose parameters to obtain a high
fidelity of the target state. In Sec.~\ref{sec:SimResults}, we numerically
study the effects of both imperfect control pulses and the environment on the
generated target state. In Sec.~\ref{sec:discussions}, the advantages and
experimental feasibility of our method are discussed. Finally, we summarize
our results in Sec.~\ref{sec:conclusions}.

\begin{figure}[ptb]
\includegraphics[width=0.46\textwidth, clip]{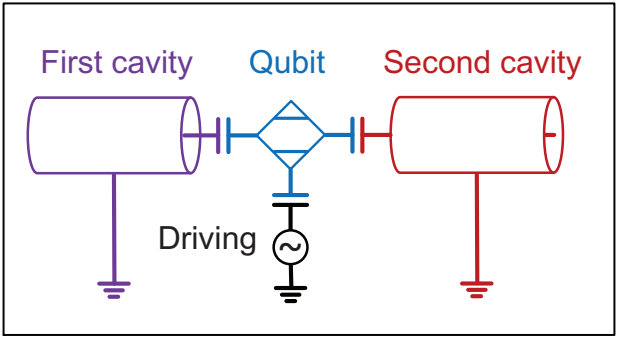}\caption{(Color online)
Schematic diagram for a driven qubit (in the middle with the blue color),
which is coupled to two single-mode microwave fields of two separated cavities
(in the left with the purple color and the right with the red color,
respectively). The first cavity field has the frequency $\omega_{1}$ and the
second one has the frequency $\omega_{2}$. The coupling strength is $g_{1}$
($g_{2}$) between the qubit and the first (second) cavity field. The qubit is
driven by a classical field (in the middle with the black color) with the
frequency $\tilde{\omega}$ and Rabi frequency $\Omega$.}%
\label{fig:model}%
\end{figure}

\section{Theoretical model and sideband excitations}

\label{sec:inter-mode}

\subsection{Basic Hamiltonian}

As schematically shown in Fig.~\ref{fig:model}, we study a system where a
superconducting qubit (SQ), modeled as a two level system, is coupled to two
single-mode microwave fields in two separated cavities and driven by a
classical field. The system Hamiltonian can be given by
\begin{equation}
\tilde{H}=\tilde{H}_{q}+H_{r}+\tilde{H}_{g}+\tilde{H}_{d}. \label{eq:1}%
\end{equation}
Here, $\tilde{H}_{q}$ and $H_{r}$ are the free Hamiltonians of the SQ and the
cavity fields, respectively. Moreover, $\tilde{H}_{g}$ is the interaction
Hamiltonian between the SQ and cavity fields, and $\tilde{H}_{d}$ is the
interaction Hamiltonian between the SQ and the classical field. In the qubit
basis, the qubit Hamiltonian is given by
\begin{equation}
\tilde{H}_{q}=\hbar\omega_{q}\frac{\tilde{\sigma}_{z}}{2}, \label{eq:Hq}%
\end{equation}
with $\tilde{\sigma}_{x}=\left\vert \tilde{g}\right\rangle \left\langle
\tilde{e}\right\vert +\left\vert \tilde{e}\right\rangle \left\langle \tilde
{g}\right\vert $ and $\tilde{\sigma}_{z}=$ $\left\vert \tilde{e}\right\rangle
\left\langle \tilde{e}\right\vert -\left\vert \tilde{g}\right\rangle
\left\langle \tilde{g}\right\vert $. The parameter $\omega_{q}$ is the qubit
frequency. The kets $\left\vert \tilde{g}\right\rangle $ and $\left\vert
\tilde{e}\right\rangle $ denote the ground and excited states of the qubit, respectively.

The free Hamiltonian of two cavity fields is given by
\begin{equation}
H_{r}=\sum_{l=1}^{2}\hbar\omega_{l}a_{l}^{\dag}a_{l},
\end{equation}
where $a_{l}$ ($a_{l}^{\dag})$ is the annihilation (creation) operator of the
$l$th cavity field with its frequency $\omega_{l}$ and $\omega_{1}\neq
\omega_{2}$. The interaction Hamiltonian between the qubit and two cavity
fields is
\begin{equation}
\tilde{H}_{g}=\sum_{l=1}^{2}\hbar g_{l}\left(  \tilde{\sigma}_{z}\cos
\theta-\tilde{\sigma}_{x}\sin\theta\right)  (a_{l}^{\dag}+a_{l}),
\label{eq:Hg_tilde}%
\end{equation}
where $g_{l}$ is the coupling strength between the $l$th cavity field and the
qubit, and $\theta$ is a parameter which depends on the inversion symmetry of
the qubit potential energy.

Similarly, the interaction Hamiltonian between the qubit and classical field
is given by
\begin{equation}
\tilde{H}_{d}=\hbar\Omega\left(  \tilde{\sigma}_{z}\cos\theta-\tilde{\sigma
}_{x}\sin\theta\right)  \cos(\tilde{\omega}t+\phi), \label{eq:Hd_tilde}%
\end{equation}
where $\Omega$ is the coupling strength (or Rabi frequency) between the qubit
and the driving field. The parameters $\tilde{\omega}$ and $\phi$ are the
driving frequency and driving phase, respectively.

In Eqs.~(\ref{eq:Hg_tilde}) and (\ref{eq:Hd_tilde}), when the qubit potential
energy possesses inversion symmetry, i.e., $\cos\theta=0$, there are only
transverse couplings between the qubit and cavity fields~\cite{Liu-NJP}. If
the rotating wave approximation is further made and there is no driving
($\Omega=0$), Eq.~(\ref{eq:1}) is reduced to extensively studied
Jaynes-Cummings model~\cite{Scully1997}. When the qubit potential energy
possesses a broken inversion symmetry~\cite{Liu-PRL,Liu-NJP}, i.e.,
$\cos\theta\neq0$, there are both transverse and longitudinal couplings
between the qubit and microwave fields. The broken inversion symmetry of the
qubit potential energy can be achieved when the bias charge for the charge
qubit or the bias flux for the flux qubit is tuned off the optimal
point~\cite{Liu-PRL,Liu-NJP}. But for the phase qubit, the inversion symmetry
of the potential energy is always broken \cite{Martinis2002,Lisenfeld2007}.
Here, we will study a general method and not specify a particular qubit.

We now change the qubit basis into the current basis of the flux qubit or the
charge basis of the charge qubit. This is equivalent to diagonalizing the
operator $\tilde{\sigma}_{z}\cos\theta-\tilde{\sigma}_{x}\sin\theta$. In the
new basis, the Hamiltonian in Eq.~(\ref{eq:1}) becomes
\begin{equation}
H=H_{q}+H_{r}+H_{g}+H_{d}\text{.} \label{eq:H}%
\end{equation}
Here, the Hamiltonians $H_{q}$, $H_{g}$, and $H_{d}$ are given by
\begin{align}
H_{q}  &  =\hbar\omega_{x}\frac{\sigma_{x}}{2}+\hbar\omega_{z}\frac{\sigma
_{z}}{2},\\
H_{g}  &  =\sum_{l=1}^{2}\hbar g_{l}\sigma_{z}\left(  a_{l}^{\dag}%
+a_{l}\right)  ,\\
H_{d}  &  =\hbar\Omega\sigma_{z}\cos\left(  \tilde{\omega}t+\phi\right)  ,
\end{align}
with $\sigma_{x}=\left\vert g\right\rangle \left\langle e\right\vert
+\left\vert e\right\rangle \left\langle g\right\vert $, and $\sigma_{z}=$
$\left\vert e\right\rangle \left\langle e\right\vert -\left\vert
g\right\rangle \left\langle g\right\vert $. Hereafter, the parameters
$\omega_{x}=\omega_{q}\sin\theta$ and $\omega_{z}=\omega_{q}\cos\theta$ are
called transverse and longitudinal frequencies of the qubit , respectively.
The kets $\left\vert g\right\rangle \equiv\tilde{R}_{y}\left(  -\theta\right)
\left\vert \tilde{g}\right\rangle $\textbf{ }and\textbf{ }$\left\vert
e\right\rangle \equiv\tilde{R}_{y}\left(  -\theta\right)  \left\vert \tilde
{e}\right\rangle $\textbf{ }are persistent current states of the flux qubit or
charge states of the charge qubit. Here,\textbf{ }$\tilde{R}_{y}\left(
\varphi\right)  =\exp(-i\varphi\tilde{\sigma}_{y}/2)$\textbf{ }is the rotation
operator along the $y$-axis, with\textbf{ }$\tilde{\sigma}_{y}=$\textbf{
}$-i\left\vert \tilde{e}\right\rangle \left\langle \tilde{g}\right\vert
+i\left\vert \tilde{g}\right\rangle \left\langle \tilde{e}\right\vert $. The
parameter $\omega_{z}\neq0$ results in longitudinal couplings between the
qubit and microwave fields in Eq.~(\ref{eq:1}). Below, we will show that
$\omega_{z}\neq0$ can induce two-mode multiphoton processes in the qubit, and
then use these multiphton processes to generate arbitrary superpositions of
two-mode photon states.

\subsection{Multiphoton processes and sideband excitations}

To see how the multiphoton processes can be induced by the longitudinal
coupling when $\omega_{z}\neq0$, we now apply a unitary transformation
\begin{equation}
D=\exp\left[  \sum_{l=1}^{2}\eta_{l}\frac{\sigma_{z}}{2}\left(  a_{l}^{\dag
}-a_{l}\right)  \right]  , \label{eq:D}%
\end{equation}
to the Hamiltonian in Eq.~(\ref{eq:H}). Then, we obtain an effective
Hamiltonian
\begin{align}
H_{\mathrm{eff}}=  &  DHD^{\dag}=\hbar\omega_{z}\frac{\sigma_{z}}{2}%
+\hbar\Omega\sigma_{z}\cos\left(  \tilde{\omega}t+\phi\right)  \label{eq:Heff}%
\\
&  +\sum_{l=1}^{2}\hbar\omega_{l}a_{l}^{\dag}a_{l}+\frac{\hbar\omega_{x}}%
{2}\left[  \sigma_{+}e^{\sum_{l=1}^{2}\eta_{l}\left(  a_{l}^{\dag}%
-a_{l}\right)  }+\mathrm{H.c.}\right]  .\nonumber
\end{align}
It is clear that $D$ is the displacement operator~\cite{Scully1997} of
two-mode cavity fields. The displacement quantity is $\eta_{l}\sigma_{z}/2$
for the $l$th cavity field. Hereafter, we will call the picture after the
operator $D$ as the displacement picture. The ratios $\eta_{l}=2g_{l}%
/\omega_{l}$ are called the Lamb-Dicke parameters in analogy to trapped
ions~\cite{trappedions,wei}.

To understand the classical-field-assisted multiphoton transitions of two
cavity fields in the qubit, we apply to Eq.~(\ref{eq:Heff}) a time-dependent
unitary transformation
\begin{equation}
U_{d}\left(  t\right)  =\exp\left[  ix\frac{\sigma_{z}}{2}\sin\left(
\tilde{\omega}t+\phi\right)  \right]  , \label{eq:Ud}%
\end{equation}
with $x=2\Omega/\tilde{\omega}$. Then, another effective Hamiltonian
\begin{align}
H_{\mathrm{eff}}^{\left(  d\right)  }=  &  U_{d}H_{\mathrm{eff}}U_{d}^{\dag
}-i\hbar U_{d}\frac{\partial}{\partial t}U_{d}^{\dag}\nonumber\\
=  &  \frac{\hbar}{2}\omega_{z}\sigma_{z}+\sum_{l=1}^{2}\hbar\omega_{l}%
a_{l}^{\dag}a_{l}\nonumber\\
&  +\frac{\hbar\omega_{x}}{2}\sum_{N=-\infty}^{\infty}\left[  J_{N}\sigma
_{+}B_{N}\left(  t\right)  +\mathrm{H.c.}\right]  , \label{eq:Heffd}%
\end{align}
can be derived, with the time-dependent term
\begin{equation}
B_{N}\left(  t\right)  =\exp\left[  \sum_{l=1}^{2}\eta_{l}\left(  a_{l}^{\dag
}-a_{l}\right)  +iN\left(  \tilde{\omega}t+\phi\right)  \right]  .
\end{equation}
Here, $J_{N}\equiv J_{N}\left(  x\right)  $ is the Bessel function of the
first kind. Equation~(\ref{eq:Heffd}\textbf{)} shows that multiphoton
transitions with different modes can be controlled by the classical field as
in trapped ions~\cite{Steinbach1997}.

In the interaction picture with the free Hamiltonian $H_{0}=\sum_{l=1}%
^{2}\hbar a_{l}^{\dagger}a_{l}+(\hbar\omega_{z}\sigma_{z}/2)$,
Equation~(\ref{eq:Heffd}) becomes
\begin{align}
H_{\mathrm{int}}=  &  \frac{\hbar\omega_{x}}{2}\!\sum_{N=-\infty}^{\infty}%
\sum_{\substack{m_{1}=0\\n_{1}=0}}^{\substack{\infty\\\infty}}\sum
_{\substack{m_{2}=0\\n_{2}=0}}^{\substack{\infty\\\infty}}J_{Nm_{1}n_{1}%
}^{m_{2}n_{2}}a_{1}^{\dag m_{1}}a_{1}^{n_{1}}a_{2}^{\dag m_{2}}a_{2}^{n_{2}%
}\sigma_{+}\nonumber\\
&  +\text{H.c.}, \label{eq:Hint}%
\end{align}
where $J_{Nm_{1}n_{1}}^{m_{2}n_{2}}\equiv J_{Nm_{1}n_{1}}^{m_{2}n_{2}%
}\!\left(  t\right)  $ is the coupling strength between the qubit and cavity
field with each different transition process, and its algebraic form is%
\begin{align}
J_{Nm_{1}n_{1}}^{m_{2}n_{2}}\!=  &  \exp\left\{  i\left[  N\tilde{\omega
}+\omega_{z}+\sum_{l=1}^{2}\left(  m_{l}-n_{l}\right)  \omega_{l}\right]
t+iN\phi\right\} \nonumber\\
&  \times\exp\left(  -{\sum_{l}\frac{\eta_{l}^{2}}{2}}\right)  J_{N}\left(
x\right)  \frac{\left(  -1\right)  ^{n_{1}+n_{2}}\eta_{1}^{m_{1}+n_{1}}%
\eta_{2}^{m_{2}+n_{2}}}{m_{1}!n_{1}!m_{2}!n_{2}!}. \label{eq:JNab}%
\end{align}
Equation~(\ref{eq:Hint}) describes the classical-field-assisted two-mode
multiphoton processes as in trapped ions~\cite{Steinbach1997}. The magnitude
of $J_{Nm_{1}n_{1}}^{m_{2}n_{2}}$ depends on $\omega_{x}$, $x$, and $\eta_{l}%
$. We find
\begin{equation}
\left\vert J_{Nm_{1}n_{1}}^{m_{2}n_{2}}\right\vert =\left\vert J_{N}%
\right\vert \left\vert \frac{J_{N}^{m_{1}n_{1}}}{J_{N}}\right\vert \left\vert
\frac{J_{N}^{m_{2}n_{2}}}{J_{N}}\right\vert , \label{eq:JNab_ABS}%
\end{equation}
where the properties of $J_{N}^{m_{l}n_{l}}\equiv J_{N}^{m_{l}n_{l}}(t)$ have
been studied in Ref.~\cite{Zhao2015}. The specific expression\textbf{ }of
$J_{N}^{m_{l}n_{l}}$ is given by
\begin{align}
J_{N}^{m_{l}n_{l}}=  &  \frac{(-1)^{n_{l}}J_{N}\!\left(  x\right)  }%
{m_{l}!n_{l}!}\eta_{l}^{m_{l}+n_{l}}\exp\left(  -\frac{\eta_{l}^{2}}{2}\right)
\label{eq:JNmn}\\
&  \times\exp\left\{  i\left[  N\tilde{\omega}+\omega_{z}+\left(  m_{l}%
-n_{l}\right)  \omega_{l}\right]  t+iN\phi\right\}  .\nonumber
\end{align}
Similarly to Eq.~(\ref{eq:JNab_ABS}), the magnitude of $J_{N}^{m_{l}n_{l}}$
can be rewritten as
\begin{equation}
\left\vert J_{N}^{m_{l}n_{l}}\right\vert =\left\vert J_{N}\right\vert
\left\vert \frac{J_{N}^{m_{l}n_{l}}}{J_{N}}\right\vert . \label{eq:JN_ABS}%
\end{equation}
It is clear that $|J_{N}^{m_{l}n_{l}}/J_{N}|$ is independent of the reduced
driving strength $x$. From Eqs.~(\ref{eq:JNab_ABS})-(\ref{eq:JN_ABS}), we know
that both $|J_{Nm_{1}n_{1}}^{m_{2}n_{2}}|$ and $|J_{N}^{m_{l}n_{l}}|$ can be
changed by adjusting $x$ and $\eta_{l}$ in a similar way. By introducing new
variables $k_{l}=m_{l}-n_{l}$, we expand Eq.~(\ref{eq:Hint}) in the Fock state
basis, and then have
\begin{align}
H_{\mathrm{int}}=  &  \hbar\sum_{N,k_{1},k_{2}=-\infty}^{\infty}\sum
_{n_{1}=\xi_{1}}^{\infty}\sum_{n_{2}=\xi_{2}}^{\infty}W_{N\zeta_{1}\zeta_{2}%
}^{k_{1}k_{2}}\!\left(  t\right)  \sigma_{+}\sigma_{n_{1}+k_{1},n_{1}}%
^{(1)}\sigma_{n_{2}+k_{2},n_{2}}^{(2)}\nonumber\\
&  +\text{\textrm{H.c..}} \label{eq:Hint2}%
\end{align}
with $\xi_{l}=\max\left\{  0,-k_{l}\right\} $ and $\zeta_{l}=\min\left\{
n_{l},n_{l}+k_{l}\right\}  $. Here, $\sigma_{n_{l}n_{l}^{\prime}%
}^{(l)}=\left\vert n_{l}\right\rangle \left\langle n_{l}^{\prime}\right\vert $
denotes the ladder operator of the $l$th cavity field. The time-dependent
transition element $W_{N\zeta_{1}\zeta_{2}}^{k_{1}k_{2}}(t)$ is given by
\begin{equation}
W_{Nn_{1}n_{2}}^{k_{1}k_{2}}\left(  t\right)  =\Omega_{Nn_{1}n_{2}}%
^{k_{1}k_{2}}\exp(i\Delta_{N}^{k_{1}k_{2}}t),
\end{equation}
with $n_{1}$, $n_{2}$ replaced by $\zeta_{1}$, $\zeta_{2}$ respectively. The
complex transition amplitude $\Omega_{Nn_{1}n_{2}}^{k_{1}k_{2}}$ and detuning
$\Delta_{N}^{k_{1}k_{2}}$ are respectively
\begin{align}
\Omega_{Nn_{1}n_{2}}^{k_{1}k_{2}}  &  =\frac{\omega_{x}}{2}J_{N}\left(
x\right)  M_{n_{1}}^{k_{1}}(\eta_{1})M_{n_{2}}^{k_{2}}\left(  \eta_{2}\right)
e^{iN\phi},\label{eq:Rabi}\\
\Delta_{N}^{k_{1}k_{2}}  &  =N\tilde{\omega}+\omega_{z}+k_{1}\omega_{1}%
+k_{2}\omega_{2}. \label{eq:detuning}%
\end{align}
The parameter $M_{n_{l}}^{k_{l}}$ is given by
\begin{equation}
M_{n_{l}}^{k_{l}}(\eta_{l})=(-1)^{k_{l}\varepsilon_{k_{l}}}\frac{\eta
_{l}^{|k_{l}|}}{e^{\eta_{l}^{2}/2}}\sqrt{\frac{n_{l}!}{(n_{l}+|k_{l}|)!}%
}L_{n_{l}}^{(|k_{l}|)}\!(\eta_{l}^{2}),
\end{equation}
with
\begin{align}
\varepsilon_{k}  &  =\left\{
\begin{array}
[c]{cc}%
1, & k<0\\
0, & k\geq0
\end{array}
\right.  ,\\
L_{n}^{\left(  k\right)  }\left(  z\right)   &  =\sum\limits_{l=0}^{n}%
\frac{\left(  n+k\right)  !}{\left(  n-l\right)  !}\frac{\left(  -1\right)
^{l}}{\left(  l+k\right)  !}\frac{z^{l}}{l!}.
\end{align}
Here, $L_{n}^{\left(  k\right)  }\left(  z\right)  $ is the generalized
Laguerre polynomials. It is clear that the classical-field-assisted
multiphoton transitions can be derived from Eq.~(\ref{eq:Hint2}) using
different frequency-matching conditions.

\subsection{Time evolution operators}

\label{sec:Eng_inter-mode}

We now give detailed discussions on how to engineer two-mode multiphoton
processes by tuning the driving field. Let us assume that the driving field is
tuned to satisfy the resonant condition
\begin{equation}
\Delta_{N}^{k_{1}k_{2}}=N\tilde{\omega}+\omega_{z}+k_{1}\omega_{1}+k_{2}%
\omega_{2}=0. \label{eq:res_con}%
\end{equation}
Then Eq.~(\ref{eq:Hint2}) can be reduced to an effective Hamiltonian
$H_{N}^{k_{1}k_{2}}$ when unwanted terms are neglected, that is,%

\begin{align}
H_{N}^{k_{1}k_{2}}=  &  \hbar\sum_{n_{1}=\xi_{1}}^{\infty}\sum_{n_{2}=\xi_{2}%
}^{\infty}\Omega_{N\zeta_{1}\zeta_{2}}^{k_{1}k_{2}}\!\sigma_{+}\sigma
_{n_{1}+k_{1},n_{1}}^{(1)}\sigma_{n_{2}+k_{2},n_{2}}^{(2)}\nonumber\\
&  +\text{\textrm{H.c..}} \label{eq:H_sideband1}%
\end{align}
The time evolution operator governed by the Hamiltonian in
Eq.~(\ref{eq:H_sideband1}) is given by
\begin{align}
U_{N}^{k_{1}k_{2}}=  &  \sum_{n_{1}=\xi_{1}}^{\infty}\sum_{n_{2}=\xi_{2}%
}^{\infty}C_{N\zeta_{1}\zeta_{2}}^{k_{1}k_{2}}\sigma_{gg}\sigma_{n_{1}n_{1}%
}^{(1)}\sigma_{n_{2}n_{2}}^{(2)}\nonumber\\
&  -\sum_{n_{1}=\xi_{1}}^{\infty}\sum_{n_{2}=\xi_{2}}^{\infty}iS_{N\zeta
_{1}\zeta_{2}}^{k_{1}k_{2}}\sigma_{+}\sigma_{n_{1}+k_{1},n_{1}}^{(1)}%
\sigma_{n_{2}+k_{2},n_{2}}^{(2)}\nonumber\\
&  -\sum_{n_{1}=\xi_{1}}^{\infty}\sum_{n_{2}=\xi_{2}}^{\infty}iS_{N\zeta
_{1}\zeta_{2}}^{k_{1}k_{2}\ast}\sigma_{-}\sigma_{n_{1},n_{1}+k_{1}}%
^{(1)}\sigma_{n_{2},n_{2}+k_{2}}^{(2)}\nonumber\\
&  +\sum_{n_{1}=\xi_{1}}^{\infty}\sum_{n_{2}=\xi_{2}}^{\infty}C_{N\zeta
_{1}\zeta_{2}}^{k_{1}k_{2}}\sigma_{ee}\sigma_{n_{1}+k_{1}\text{,}n_{1}+k_{1}%
}^{(1)}\sigma_{n_{2}+k_{2}\text{,}n_{2}+k_{2}}^{(2)}\nonumber\\
&  +\sum_{n_{1}<-k_{1}\text{ or }n_{2}<-k_{2}}\sigma_{gg}\sigma_{n_{1}n_{1}%
}^{(1)}\sigma_{n_{2}n_{2}}^{(2)},\nonumber\\
&  +\sum_{n_{1}<k_{1}\text{ or }n_{2}<k_{2}}\sigma_{ee}\sigma_{n_{1}n_{1}%
}^{(1)}\sigma_{n_{2}n_{2}}^{(2)}, \label{eq:U_sideband1}%
\end{align}
Recall that $\xi_{l}=\max\left\{  0,-k_{l}\right\}  $ and $\zeta_{l}%
=\min\left\{  n_{l},n_{l}+k_{l}\right\}  $ as defined previously. Here the new
parameters used in Eq.~(\ref{eq:U_sideband1}) are respectively
\begin{align}
C_{Nn_{1}n_{2}}^{k_{1}k_{2}}  &  \equiv C_{Nn_{1}n_{2}}^{k_{1}k_{2}}\left(
t\right)  =\cos\left(  \left\vert \Omega_{Nn_{1}n_{2}}^{k_{1}k_{2}}\right\vert
t\right)  ,\\
S_{Nn_{1}n_{2}}^{k_{1}k_{2}}  &  \equiv S_{Nn_{1}n_{2}}^{k_{1}k_{2}}\left(
t\right)  =e^{i\phi_{Nn_{1}n_{2}}^{k_{1}k_{2}}}\sin\left(  \left\vert
\Omega_{Nn_{1}n_{2}}^{k_{1}k_{2}}\right\vert t\right)  ,\\
\phi_{Nn_{1}n_{2}}^{k_{1}k_{2}}  &  =\arg\left(  \Omega_{Nn_{1}n_{2}}%
^{k_{1}k_{2}}\right)  . \label{eq:phiNK}%
\end{align}
As shown in Eqs.~(\ref{eq:H_sideband1}) and (\ref{eq:U_sideband1}),
$\left\vert k_{l}\right\vert $ photons in the $l$th resonator can be either
created if $k_{l}\geq0$ or annihilated if $k_{l}<0$ while the qubit is flipped
up. Similarly, $\left\vert k_{l}\right\vert $ photons in the $l$th resonator
can be either created if $k_{l}<0$ or annihilated if $k_{l}\geq0$ while the
qubit is flipped down. Thus different sideband excitations can be constructed,
depending on the values of $k_{1}$ and $k_{2}$.

Because the Hamiltonian derived in Eq.~(\ref{eq:Hint}) is similar to that of
the trapped ions~\cite{Zou2002}, the algorithm using two-mode multiphonon
processes in trapped ions can be directly applied into our model, and
different superpositions of two-mode photons can be generated. As a special
case, two-mode Fock states of high photon numbers can in principle be more
efficiently generated with just two steps as single-mode Fock states of high
photon numbers~\cite{Zhao2015}. However, we here design a new algorithm via
different sideband transitions of low photon numbers by tuning the driving
field with properly selecting the parameters $\omega_{z}$, $\omega_{x}$,
$\omega_{l}$, and $\eta_{l}$. The detailed discussions of parameter selection
will be given in Sec.~\ref{sec:ParSel}.

\section{Algorithm for state generation\label{sec:algorithm}}

Let us first study a universal algorithm for generating arbitrary two-mode
microwave photon states using sideband transitions with the following four
Hamiltonians $H_{\bar{1}}^{1\bar{1}}$, $H_{\bar{1}}^{\bar{1}0}$, $H_{\bar{1}%
}^{0\bar{1}}$, and $H_{\bar{1}}^{00}$. Here, for the compact of notations, we
have used $\bar{k}$ to represent $-k$ with $k>0$. For instance, $H_{\bar{1}%
}^{1\bar{1}}$ is actually $H_{N}^{k_{1}k_{2}}$ with $N=-1$, $k_{1}=1$, and
$k_{2}=-1$. For different $N$, $k_{1}$, and $k_{2}$, the interaction
Hamiltonian $H_{N}^{k_{1}k_{2}}$ and its time evolution operator $U_{N}%
^{k_{1}k_{2}}$ have already been given in Sec.~\ref{sec:Eng_inter-mode}.
Below, we will first study how to generate the target state by choosing pulse
durations, frequencies, and phases of the driving fields at each generation
step with different sideband excitations, and then we will apply our algorithm
to the generation of NOON states and discuss particular properties of the algorithm.

\subsection{Universal algorithm for generating arbitrary two-mode microwave
photon states}

We note that the state generation in our algorithm is studied in the
displacement picture with the unitary transformation as shown in
Eq.~(\ref{eq:D}). The arbitrary quantum states, we expect to be generated, is
written as%
\begin{equation}
\left\vert \psi_{f}\right\rangle =\sum_{n_{1}+n_{2}\leq N_{\max}}C_{n_{1}%
n_{2}}\left\vert n_{1},n_{2}\right\rangle \left\vert g\right\rangle ,
\label{eq:psi_f}%
\end{equation}
where $\left\vert n_{1},n_{2}\right\rangle $ means that the first and second
cavities contain $n_{l}$ and $n_{2}$ photons,respectively, and $\left\vert
g\right\rangle $ means that the qubit is in the ground state. Besides,
$N_{\max}$ and $C_{n_{1}n_{2}}$ mean the maximum photon number and the
probability amplitude on the state $\left\vert n_{1},n_{2}\right\rangle
\left\vert g\right\rangle $, respectively. We assume that the system is
initially in the state
\begin{equation}
\left\vert \psi_{0}\right\rangle =\left\vert 0,0\right\rangle \left\vert
g\right\rangle . \label{eq:psi_0}%
\end{equation}

We suppose the target state $\left\vert \psi_{f}\right\rangle $ can be
generated by alternately switching on and off the two-mode transitions
$H_{\bar{1}}^{1\bar{1}}$, $H_{\bar{1}}^{\bar{1}0}$, $H_{\bar{1}}^{0\bar{1}}$,
and $H_{\bar{1}}^{00}$. With the designed time evolution operators, the state
generation procedure can be represented by,%
\begin{align}
\left\vert \psi_{f}\right\rangle =  &  \prod\limits_{\nu=f}^{1\text{ }}\left[
\bar{U}_{\nu}^{\dag}\left(  t_{\nu}\right)  U_{\bar{1}}^{p_{\nu}}\left(
t_{\nu}\right)  \bar{U}_{_{\nu}}\left(  0\right)  \right]  \left\vert \psi
_{0}\right\rangle \nonumber\\
=  &  \left[  \bar{U}_{f}^{\dag}\left(  t_{f}\right)  U_{\bar{1}}^{p_{f}%
}\left(  t_{f}\right)  \bar{U}_{f}\left(  0\right)  \right]  \cdots\nonumber\\
&  \left[  \bar{U}_{1}^{\dag}\left(  t_{1}\right)  U_{\bar{1}}^{p_{1}}\left(
t_{1}\right)  \bar{U}_{1}\left(  0\right)  \right]  \left\vert
0,0\right\rangle \left\vert g\right\rangle , \label{eq:evolution}%
\end{align}
where $p_{\nu}\in\left\{  1\bar{1},\bar{1}0,0\bar{1},00\right\}  $ denotes the
transition type for the $\nu$th step, and $t_{\nu}$ is the time duration for
the $\nu$th step. The time evolution operator $\bar{U}_{\nu}\left(  t_{\nu
}\right)  $ is given by
\begin{align}
\bar{U}_{\nu}\left(  t_{\nu}\right)  =  &  \exp\left[  i\left(  \omega
_{z}\frac{\sigma_{z}}{2}+\omega_{1}a_{1}^{\dag}a_{1}+\omega_{2}a_{2}^{\dag
}a_{2}\right)  t_{\nu}\right] \nonumber\\
&  \times\exp\left[  ix\frac{\sigma_{z}}{2}\sin\left(  \tilde{\omega}_{\nu
}t_{\nu}+\phi_{\nu}\right)  \right]  . \label{eq:Ubar}%
\end{align}
As discussed above, the transitions of different types can be achieved by
changing the frequency $\tilde{\omega}$ of the driving field, which is denoted
by $\tilde{\omega}_{\nu}$ for the $\nu$th step. The phase of the driving field
for the $\nu$th step is denoted by $\phi_{\nu}$. We can express
Eq.~(\ref{eq:evolution}) in another equivalent form of iteration,
\begin{equation}
\left\vert \psi_{\nu-1}\right\rangle =\bar{U}_{\nu}^{\dag}\left(  0\right)
U_{\bar{1}}^{p_{\nu}\dag}\left(  t_{\nu}\right)  \bar{U}_{\nu}\left(  t_{\nu
}\right)  \left\vert \psi_{\nu}\right\rangle , \label{eq:iteration}%
\end{equation}
with $|\psi_{0}\rangle$ and $|\psi_{f}\rangle$ given in Eqs.~(\ref{eq:psi_0})
and (\ref{eq:psi_f}), respectively. The ket $|\psi_{\nu}\rangle$ is the state
after the $\nu$th step. We note that the subscript $f$ of $|\psi_{f}\rangle$
in Eq.~(\ref{eq:evolution}) denotes the number of the final step.
Equation~(\ref{eq:iteration}) means that the initial state is restored from
the target state by a composition of sideband transitions with proper time
durations, frequencies and phases of driving fields. It is a recursion algorithm.

\begin{figure}[ptbh]
\includegraphics[width=0.48\textwidth, clip]{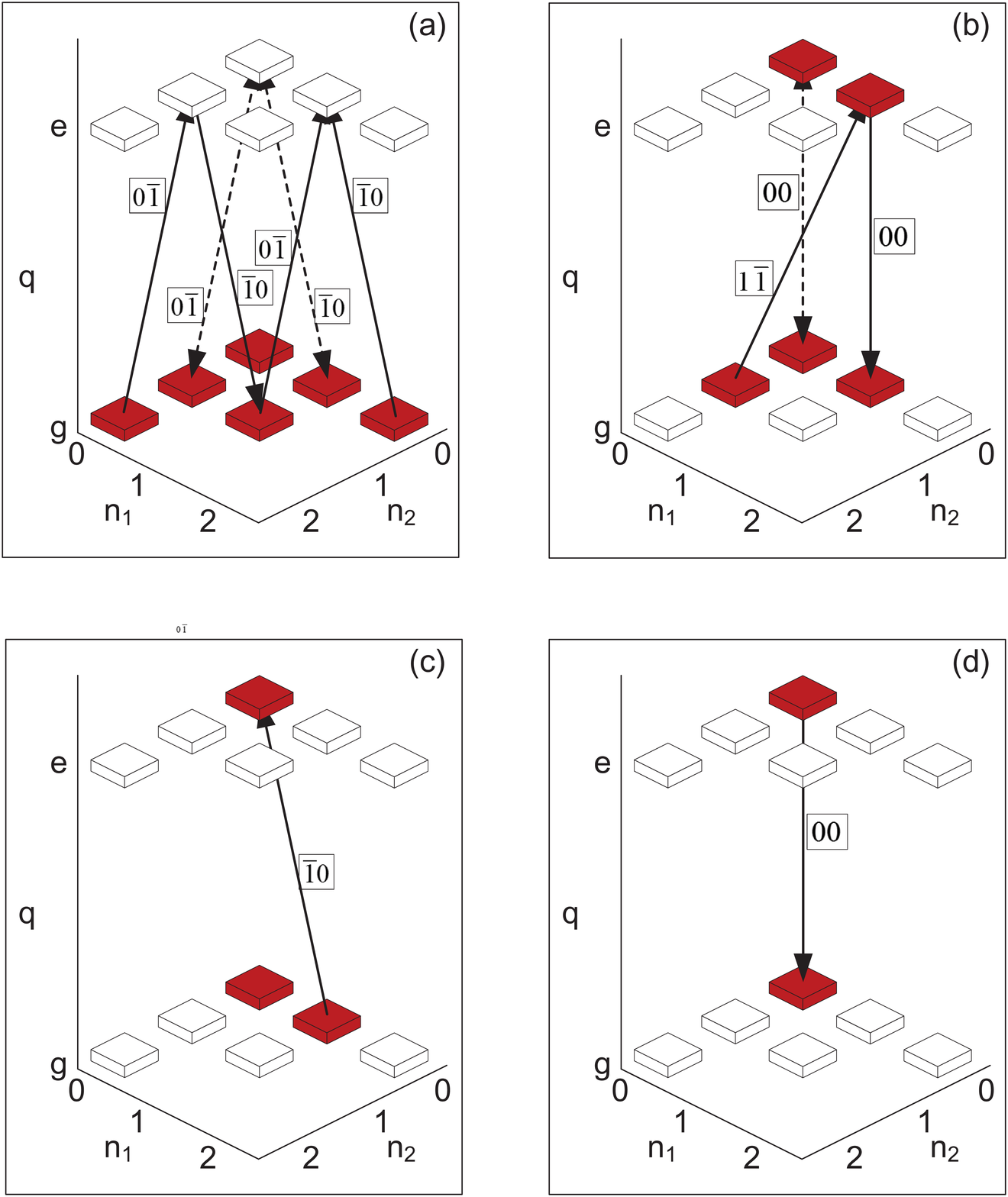} \caption{(Color online)
Universal algorithm for generating arbitrary two-mode superposition state with
the maximal photon number $N_{\max}=2$. The $n_{1}$ and $n_{2}$-axis
respectively denote the photon number of the first and second mode. The
two-mode photon state is denoted by $|n_{1},n_{2}\rangle$. The qubit state is
represented by the $q$-axis with $q=g$ or $e$ respectively denoting the ground
state $|g\rangle$ or excited state $|e\rangle$. The state component
$|n_{1},n_{2}\rangle|q\rangle$ is represented by a block at the location
$(n_{1},n_{2},q)$. If a state component is occupied, we color the
corresponding block with red; otherwise, the block is left uncolored. The
arrows respectively represent the \textquotedblleft\={1}0 \textquotedblright,
\textquotedblleft0\={1}\textquotedblright, \textquotedblleft1\={1}%
\textquotedblright, and \textquotedblleft00\textquotedblright\ transitions
with transition types labeled aside them. The solid arrow indicates a desired
population transfer from the starting state to the end state, while the dashed
arrow indicates the inevitable irrelevant oscillation when the desired
population transfer is implemented. (a) Schematic diagram for transferring the
populations on states $|0,2\rangle|g\rangle$, $|1,1\rangle|g\rangle$, and
$|2,0\rangle|g\rangle$ to the state $|1,0\rangle|e\rangle$. This is achieved
by consecutively using \textquotedblleft0\={1}\textquotedblright,
\textquotedblleft\={1}0\textquotedblright, \textquotedblleft0\={1}%
\textquotedblright, and \textquotedblleft\={1}0\textquotedblright%
\ transitions. (b) Schematic diagram for transferring populations on states
$|0,1\rangle|g\rangle$ and $|1,0\rangle|e\rangle$ to the state $|1,0\rangle
|g\rangle$. This is achieved by consecutively using \textquotedblleft%
1\={1}\textquotedblright\ and \textquotedblleft00\textquotedblright%
\ transitions. (c) Schematic diagram for transferring the population on the
state $|1,0\rangle|g\rangle$ to the state $|0,0\rangle|e\rangle$. This is
achieved by using a \textquotedblleft\={1}0\textquotedblright\ transition. (d)
Schematic diagram for transferring the\ population on the state $|0,0\rangle
|e\rangle$ to the state $|0,0\rangle|g\rangle$. This is achieved by using a
\textquotedblleft00\textquotedblright\ transition. }%
\label{fig:general}%
\end{figure}

Without loss of generality, we use the maximum photon number $N_{\max}=2$ as
an example to show our algorithm. The more general case with arbitrary
$N_{\max}$ is given in Appendix.~\ref{append:Univ_alg}. The detailed steps for
generating the target state%
\begin{align}
|\psi_{f}\rangle=  &  C_{02}\left\vert 0,2\right\rangle \left\vert
g\right\rangle +C_{11}\left\vert 1,1\right\rangle \left\vert g\right\rangle
+C_{20}\left\vert 2,0\right\rangle \left\vert g\right\rangle \nonumber\\
+ & C_{01}\left\vert 0,1\right\rangle \left\vert g\right\rangle +C_{10}%
\left\vert 1,0\right\rangle \left\vert g\right\rangle +C_{00}\left\vert
0,0\right\rangle \left\vert g\right\rangle ,
\end{align}
with $N_{\max}=2$ using our recursion algorithm are described as the following
four procedures.

\textbf{Procedure (i)}. As schematically shown in Fig.~\ref{fig:general}(a),
from the final state $|\psi_{f}\rangle$, we first transfer the populations in
the state space spanned by $\{|n_{1},n_{2}\rangle|g\rangle|n_{1}+n_{2}=2\}$ to
the state $|1,0\rangle|e\rangle$. This procedure consists of four
steps as schematically shown in below
\begin{align}
&  \left\vert 0,2\right\rangle \left\vert g\right\rangle
\xrightarrow[f]{0\bar{1}}\left\vert 0,1\right\rangle \left\vert e\right\rangle
\xrightarrow[f-1]{\bar{1}0}\left\vert 1,1\right\rangle \left\vert
g\right\rangle \xrightarrow[f-2]{0\bar{1}}\nonumber\\
&  \left\vert 1,0\right\rangle \left\vert e\right\rangle
\xleftarrow[f-3]{\bar{1}0}\left\vert 2,0\right\rangle \left\vert
g\right\rangle .\label{eq:Proc_i}%
\end{align}
In Eq.~(\ref{eq:Proc_i}), the transition type and step number is labeled
respectively above and below the arrow. The arrow points to the direction of
the population transfer. In Step $\nu$, the population transfer is
accomplished by properly tuning $\tilde{\omega}_{\nu}$, $t_{\nu}$, and
$\phi_{\nu}$. After this procedure, we obtain the state $|\psi_{f-4}\rangle$
which only has popupaltions in the space $\{\left\vert 0,0\right\rangle
\left\vert g\right\rangle ,\left\vert 0,0\right\rangle \left\vert
e\right\rangle ,\left\vert 0,1\right\rangle \left\vert g\right\rangle
,\left\vert 1,0\right\rangle \left\vert g\right\rangle ,\left\vert
1,0\right\rangle \left\vert e\right\rangle \}$. We note that two additional
oscillations
\begin{equation}
|1,1\rangle|g\rangle\overset{0\bar{1}}{\leftrightarrow}|1,0\rangle
|e\rangle\text{ and }|0,1\rangle|g\rangle\overset{0\bar{1}}{\leftrightarrow
}|0,0\rangle|e\rangle.
\end{equation}
will also occur inevitably when the population transfer from the state
$\left\vert 0,2\right\rangle \left\vert g\right\rangle $ to the state
$\left\vert 0,1\right\rangle \left\vert e\right\rangle $ is implemented. But
they do not cause population leakage outside the original space and no extra steps should be taken for them. We call these oscillations as
irrelevant oscillations. For this procedure, these irrelevant oscillations are
schematically shown by dashed arrows in Fig.~\ref{fig:general}(a). The
irrelevant oscillation can also occur in the following procedures and is shown by
dashed arrows.

\textbf{Procedure (ii)}. As schematically shown in Fig.~\ref{fig:general}(b),
starting from the state $|\psi_{f-4}\rangle$, we need to transfer the
populations in the state space spanned by $\{|0,1\rangle|g\rangle
,|1,0\rangle|e\rangle\}$ to the state $|1,0\rangle|g\rangle$. This
procedure consists of following two steps
\begin{equation}
\left\vert 0,1\right\rangle \left\vert g\right\rangle
\xrightarrow[f-4]{1\bar{1}}\left\vert 1,0\right\rangle \left\vert
e\right\rangle \xrightarrow[f-5]{00}\left\vert 1,0\right\rangle \left\vert
g\right\rangle .
\end{equation}
After this procedure, we obtain the state $|\psi_{f-6}\rangle$, which only has
popupaltions in the space $\{\left\vert 0,0\right\rangle \left\vert
g\right\rangle ,\left\vert 0,0\right\rangle \left\vert e\right\rangle
,\left\vert 1,0\right\rangle \left\vert g\right\rangle \}$.

\textbf{Procedure (iii)}. This procedure is similar to Procedure (i). As
schematically shown in Fig.~\ref{fig:general}(c), starting from the state
$|\psi_{f-6}\rangle$, here we need to transfer the population on the state
$|1,0\rangle|g\rangle$ to the state $|0,0\rangle|e\rangle$. This
procedure consists of only one step as below
\begin{equation}
\left\vert 1,0\right\rangle \left\vert g\right\rangle
\xrightarrow[f-6]{\bar{1}0}\left\vert 0,0\right\rangle \left\vert
e\right\rangle .
\end{equation}
After this procedure, we obtain the state $|\psi_{f-7}\rangle$, which only has
popupaltions in the space $\{\left\vert 0,0\right\rangle \left\vert
g\right\rangle ,\left\vert 0,0\right\rangle \left\vert e\right\rangle \}$.

\textbf{Procedure (iv)}. This procedure is similar to the Procedure (ii). As
schematically shown in Fig.~\ref{fig:general}(d), starting from the state
$|\psi_{f-7}\rangle$, we need to transfer the population on the state
$|0,0\rangle|e\rangle$ to the state $|0,0\rangle|g\rangle$. This
procedure only consists of one step as below
\begin{equation}
\left\vert 0,0\right\rangle \left\vert e\right\rangle
\xrightarrow[f-7]{00}\left\vert 0,0\right\rangle \left\vert g\right\rangle .
\end{equation}
We thus obtain the state $|\psi_{f-8}\rangle=|0,0\rangle|g\rangle$.

Therefore, the target $\left\vert \psi_{f}\right\rangle $ can be generated
from the initial state $\left\vert \psi_{f-8}\right\rangle $ using inverse
processes from the Procedure (iv) to the Procedure (i). We note $\left\vert
\psi_{f-8}\right\rangle \equiv\left\vert \psi_{0}\right\rangle =|0,0\rangle
|g\rangle$. Thus, we obtain the total step number $f=8$ by setting $f-8=0$.
Therefore, the generation of the target state with $N_{\mathrm{max}}=2$ needs
$8$ steps.

Our algorithm takes a quadratic number of steps while an exponential one is
required in Ref.~\cite{Gardiner1997}. Let us now analyze the reason. Our
algorithm employs four interaction Hamiltonians, $H_{\bar{1}}^{\bar{1}0}$,
$H_{\bar{1}}^{0\bar{1}}$, $H_{\bar{1}}^{00}$, and $H_{\bar{1}}^{1\bar{1}}$
given in Eq.~(\ref{eq:H_sideband1}). However, four interaction Hamiltonians
\textquotedblleft$\hat{a}_{x}\sigma^{+}+\mathrm{H.c.}$\textquotedblright,
\textquotedblleft$\hat{a}_{y}^{\dag}\sigma^{-}+\mathrm{H.c.}$%
\textquotedblright, \textquotedblleft$\sigma^{-}+\sigma^{+}$\textquotedblright%
, and \textquotedblleft$\hat{a}_{x}\hat{a}_{y}\sigma^{+}+\mathrm{H.c.}%
$\textquotedblright are employed in Ref.~\cite{Gardiner1997}. The former three
interaction Hamiltonians between our algorithm and those in
Ref.~\cite{Gardiner1997} are qualitatively identical since they convert the
same number of bosons for either mode when the two-level system is excited.
However, the last ones show fundamental difference between our algorithm and
that in Ref.~\cite{Gardiner1997}, because ours creates one boson (photon) of
one mode but annihilate one boson (photon) of the other when the two-level
system is excited. But in Ref.~\cite{Gardiner1997}, one boson for both modes
can be simultaneously created when the two-level system is excited. This
difference is critical for us to design an algorithm which can keep track of
the populations with a constant total boson (photon) number. Therefore, there
is no population leakage outside the original space. However, the
algorithm in Ref.~\cite{Gardiner1997} has population leakage. Obviously, if
the last interaction in Ref.~\cite{Gardiner1997} is changed to
\textquotedblleft$\hat{a}_{x}^{\dag}\hat{a}_{y}\sigma^{+}+H.c.$%
\textquotedblright, a theoretically equivalent algorithm to ours can also be
developed. In this sense, our algorithm can be regarded as the improved
version of that in Ref.~\cite{Gardiner1997}.

\subsection{Calculation of controllable parameters}

Let us now study how to choose the pulse duration $t_{\nu}$, the frequency
$\tilde{\omega}_{\nu}$ and phase $\phi_{\nu}$ of the driving field to generate
a target state in the $\nu$th step for different types of transitions.

We suppose that the population transfer is taken as following
\begin{equation}
\left\vert n_{1},n_{2}\right\rangle \left\vert g\right\rangle
\xrightarrow[\nu]{p_{\nu}}\left\vert n_{1}+k_{1},n_{2}+k_{2}\right\rangle
\left\vert e\right\rangle ,
\end{equation}
in the $\nu$th step, where the transition type $p_{\nu}=k_{1}k_{2}$ should be
switched on based on the previous discussions. Thus the driving frequency is
taken as
\begin{equation}
\tilde{\omega}_{\nu}=\omega_{z}+k_{1}\omega_{1}+k_{2}\omega_{2},
\label{eq:omega_nu}%
\end{equation}
from the resonant condition in Eq.~(\ref{eq:res_con}). By introducing the
notations%
\begin{align}
C_{n_{1}n_{2}}^{\left(  \nu\right)  }  &  =\left\langle n_{1},n_{2}\right\vert
\left\langle g\right\vert |\psi_{\nu}\rangle,\\
D_{n_{1}n_{2}}^{\left(  \nu\right)  }  &  =\left\langle n_{1},n_{2}\right\vert
\left\langle e\right\vert |\psi_{\nu}\rangle,
\end{align}
then from\ Eq.~(\ref{eq:iteration}), we need to solve the equation,%
\begin{equation}
\left\langle n_{1},n_{2}\right\vert \left\langle g\right\vert \bar{U}_{\nu
}^{\dag}\left(  0\right)  U_{\bar{1}}^{p_{\nu}\dag}\left(  t_{\nu}\right)
\bar{U}_{\nu}\left(  t_{\nu}\right)  \left\vert \psi_{\nu}\right\rangle =0.
\end{equation}
We thus have the explicit solution for the pulse duration $t_{\nu}$ as
\begin{equation}
t_{\nu}=\frac{1}{\left\vert \Omega_{\bar{1}\zeta_{1}\zeta_{2}}^{k_{1}k_{2}%
}\right\vert }\arctan\left\vert \frac{C_{n_{1},n_{2}}^{\left(  \nu\right)  }%
}{D_{n_{1}+k_{1},n_{2}+k_{2}}^{\left(  \nu\right)  }}\right\vert .
\end{equation}
The phase of the driving field is determined by
\begin{align}
\arg\left(  \frac{C_{n_{1},n_{2}}^{\left(  \nu\right)  }}{D_{n_{1}+k_{1}%
,n_{2}+k_{2}}^{\left(  \nu\right)  }}\right)  =  &  x\sin\left(  \tilde
{\omega}_{\nu}t_{\nu}+\phi_{\nu}\right)  +\tilde{\omega}_{\nu}t_{\nu
}\nonumber\\
- & \phi_{\bar{1}\zeta_{1}\zeta_{2}}^{k_{1}k_{2}\nu}-\frac{\pi}{2}%
\operatorname{mod}2\pi.
\end{align}
Here, the notation $\phi_{\bar{1}n_{1},n_{2}}^{k_{1}k_{2}\nu}$ is the value of
$\phi_{\bar{1}n_{1},n_{2}}^{k_{1}k_{2}}$ for the $\nu$th step, which is given
in Eq.~(\ref{eq:phiNK}) and depends on $\phi_{\nu}$. Still recall $\zeta
_{l}=\min\left\{  n_{l},n_{l}+k_{l}\right\}  $ with $l=1,2$.

Similarly, if the population transfer is taken as
\begin{equation}
\left\vert n_{1}+k_{1},n_{2}+k_{2}\right\rangle \left\vert e\right\rangle
\xrightarrow[\nu]{p_{\nu}}\left\vert n_{1},n_{2}\right\rangle \left\vert
g\right\rangle .
\end{equation}
in the $\nu$th step. The explicit solution for $t_{\nu}$ is then%
\begin{equation}
t_{\nu}=\frac{1}{\left\vert \Omega_{\bar{1}\zeta_{1}\zeta_{2}}^{k_{1}k_{2}%
}\right\vert }\arctan\left\vert \frac{D_{n_{1}+k_{1},n_{2}+k_{2}}^{\left(
\nu\right)  }}{C_{n_{1},n_{2}}^{\left(  \nu\right)  }}\right\vert ,
\label{eq:t_nu_e}%
\end{equation}
and the phase of the driving field is determined by
\begin{align}
\arg\left(  \frac{C_{n_{1},n_{2}}^{\left(  \nu\right)  }}{D_{n_{1}+k_{1}%
,n_{2}+k_{2}}^{\left(  \nu\right)  }}\right)  =  &  x\sin\left(  \tilde
{\omega}_{\nu}t_{\nu}+\phi_{\nu}\right)  +\tilde{\omega}_{\nu}t_{\nu
}\nonumber\\
-  &  \phi_{\bar{1}\zeta_{1}\zeta_{2}}^{k_{1}k_{2}\nu}+\frac{\pi}%
{2}\operatorname{mod}2\pi. \label{eq:phi_nu_e}%
\end{align}

According to the target state, the time duration, frequency and
phase of the driving field for each step can be calculated using above
equations. For example, if the \textquotedblleft$00$" transition is used in
the $3$rd step, then we use Eq.~(\ref{eq:t_nu_e}) and Eq.~(\ref{eq:phi_nu_e})
to obtain $t_{3}$ and $\phi_{3}$ by setting $\nu=3$.

\subsection{Application to NOON states}

\begin{figure}[ptbh]
\includegraphics[width=0.48\textwidth, clip]{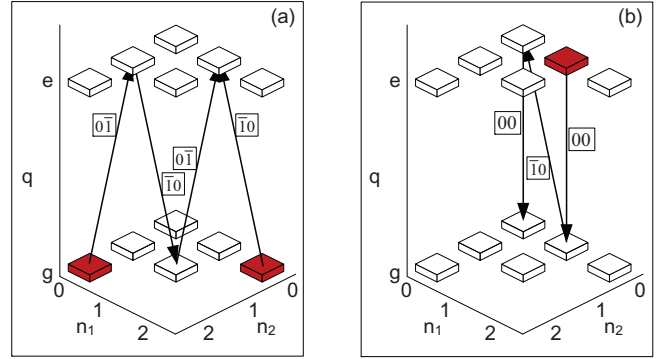}
\caption{(Color online) Application of the general algorithm to generating the
NOON state. The notations are the same as those in Fig.~\ref{fig:general}. (a)
Schematic diagram for transferring the population in the space $\{|0,2\rangle
|g\rangle,|2,0\rangle|g\rangle\}$ to the state $|1,0\rangle|e\rangle$. This is
achieved by consecutively using \textquotedblleft0\={1}\textquotedblright,
\textquotedblleft\={1}0\textquotedblright, \textquotedblleft0\={1}%
\textquotedblright,\ and \textquotedblleft\={1}0\textquotedblright%
\ transitions. (b) Schematic diagram for transferring the population on the
state $|1,0\rangle|e\rangle$ to the state $|0,0\rangle|g\rangle$. This is
achieved by consecutively using \textquotedblleft00\textquotedblright,
\textquotedblleft\={1}0\textquotedblright,\ and \textquotedblleft%
00\textquotedblright\ transitions. }%
\label{fig:noon}%
\end{figure}

As an example, we now apply our algorithm to the generation of the NOON state,
i.e., the target state is
\begin{equation}
\left\vert \psi_{f}\right\rangle =\frac{1}{\sqrt{2}}\left(  \left\vert
N_{\max},0\right\rangle \left\vert g\right\rangle +\left\vert 0,N_{\max
}\right\rangle \left\vert g\right\rangle \right)  . \label{eq:NOON}%
\end{equation}
The recursion algorithm restoring $\left\vert \psi_{f}\right\rangle $ to the
vacuum state $|0,0\rangle|g\rangle$ is schematically shown in
Fig.~\ref{fig:noon} for the maximum photon number $N_{\max}=2$. In
Fig.~\ref{fig:noon}(a), we can find that all the populations in the Hilbert
space spanned by $\{|0,2\rangle|g\rangle,|2,0\rangle|g\rangle\}$ can be
transferred to the state $|1,0\rangle|e\rangle$ by consecutively using
transitions \textquotedblleft0\={1}", \textquotedblleft\={1}0",
\textquotedblleft0\={1}", and \textquotedblleft\={1}0", i.e.,
\begin{align}
&  \left\vert 0,2\right\rangle \left\vert g\right\rangle
\xrightarrow[f]{0\bar{1}}\left\vert 0,1\right\rangle \left\vert e\right\rangle
\xrightarrow[f-1]{\bar{1}0}\left\vert 1,1\right\rangle \left\vert
g\right\rangle \xrightarrow[f-2]{0\bar{1}}\nonumber\\
&  \left\vert 1,0\right\rangle \left\vert e\right\rangle
\xleftarrow[f-3]{\bar{1}0}\left\vert 2,0\right\rangle \left\vert
g\right\rangle .
\end{align}
After this procedure, as schematically shown in
Fig.~\ref{fig:noon}(b), all the populations on the state $|1,0\rangle
|e\rangle$ can be transferred to the state $|0,0\rangle|g\rangle$ by
consecutively using transitions \textquotedblleft00", \textquotedblleft%
\={1}0", and \textquotedblleft00", i.e.,
\begin{align}
&  \left\vert 1,0\right\rangle \left\vert e\right\rangle
\xrightarrow[f-4]{00}\left\vert 1,0\right\rangle \left\vert g\right\rangle
\xrightarrow[f-5]{\bar{1}0}\nonumber\\
&  \left\vert 0,0\right\rangle \left\vert e\right\rangle
\xrightarrow[f-6]{00}\left\vert 0,0\right\rangle \left\vert g\right\rangle .
\end{align}
The total step number is thus $f=7$ for generating the NOON state
$(|0,2\rangle+|2,0\rangle)/\sqrt{2}$.

More generally, given an arbitrary $N_{\max}$, the total step number for
generating the NOON state in Eq.~(\ref{eq:NOON}) is
\begin{equation}
f=4N_{\max}-1.
\end{equation}
The step number for generating NOON state has been greatly reduced in
comparison with that for generating an arbitrary state in
Eq.~(\ref{eq:f_general}). Obviously, the NOON state can be generated without
using the \textquotedblleft1\={1}\textquotedblright\ transition. If we assume
the Lamb-Dicke parameter $\eta_{l}\lesssim1$, which is usually the case even
in the ultrastrong regime in superconducting circuit QED
systems~\cite{Niemczyk2010,Forn2010,Stassi2013}. From Eq.~(\ref{eq:Rabi}), we
have the Rabi frequencies $|\Omega_{\bar{1}n_{1}n_{2}}^{00}|\propto\eta
_{1}^{0}\eta_{2}^{0}$, $|\Omega_{\bar{1}n_{1}n_{2}}^{1\bar{1}}|\propto\eta
_{1}\eta_{2}$, $|\Omega_{\bar{1}n_{1}n_{2}}^{\bar{1}0}|\propto\eta_{1}$, and
$|\Omega_{\bar{1}n_{1}n_{2}}^{0\bar{1}}|\propto\eta_{2}$. Thus, the transition
\textquotedblleft1\={1}\textquotedblright\ generally takes more time among the
four types of transitions employed by us. Therefore, our algorithm may show a
better efficiency for generating NOON sates than generating arbitrary
entangled states. This is especially true when the maximal photon number
$N_{\max}$ is higher and the Lamb-Dicke parameter $\eta_{l}$ is smaller.

\section{Minimizing the effect of unwanted terms\label{sec:ParSel}}

\subsection{Theoretical analysis}

In all of the above studies, we make an approximation that all unwanted terms
have been neglected. However, these neglected terms will affect the fidelity
of the prepared target state. Let us now discuss how to minimize the effect of
these unwanted terms in Eq.~(\ref{eq:Hint2}) on the target state by choosing
appropriate parameters. In principle, the effects of these unwanted terms can
be perfectly removed by pulse calibration techniques. Here, we study a method
to minimize the effect of these unwanted terms by choosing the parameters when
the pulse calibration cannot be used.

In our algorithm, we have used four interactions $H_{\bar{1}}^{\bar{1}0}$,
$H_{\bar{1}}^{0\bar{1}}$, $H_{\bar{1}}^{1\bar{1}}$, and $H_{\bar{1}}^{00}$,
all of them are constructed by the terms with the Bessel function $J_{\bar{1}%
}\left(  x\right)  $ in Eq.~(\ref{eq:Hint2}). Here, in the subscript of the
Bessel function, we also use $\bar{N}$ to denote $-N$ if $N>0$. We hope to
suppress all the terms with the Bessel functions $J_{N^{\prime}}\left(
x\right)  $ for $N^{\prime}\neq-1$. We focus on the case\textbf{ }%
$x=2\Omega/\tilde{\omega}\sim1$ considering possible experimental conditions.
In this case, only lower order Bessel functions $J_{0}\left(  x\right)  $,
$J_{\pm1}\left(  x\right)  $, and $J_{\pm2}\left(  x\right)  $ play
significant roles. Thus, we need only to find proper parameters such that the
effect of the terms with $J_{0}\left(  x\right)  $, $J_{1}\left(  x\right)  $,
and $J_{\pm2}\left(  x\right)  $ are negligibly small. Our idea is to make
those terms nonresonant by properly choosing the parameters $\omega_{x}$ and
$\omega_{z}$ of the qubit, and frequencies $\omega_{1}$ and $\omega_{2}$ of
two microwave modes. That is, we assume that the frequency of the $l$th cavity
mode satisfies
\begin{equation}
\omega_{l}=l_{l}\,\omega_{\mathrm{gcd}}, \label{eq:Oil}%
\end{equation}
where $l_{l}$ is a positive integer and $\omega_{\mathrm{\gcd}}$ is the
greatest common divisor of $\omega_{1}$ and $\omega_{2}$. Assuming that the
\textquotedblleft$k_{1}k_{2}$\textquotedblright transition is switched on,
i.e., the transition detuning $\Delta_{\bar{1}}^{k_{1}k_{2}}=0$, then from
Eq.~(\ref{eq:detuning}), the frequency $\tilde{\omega}$ of the driving field
must satisfy the condition
\begin{equation}
\tilde{\omega}=\omega_{z}+k_{1}\omega_{1}+k_{2}\omega_{2}.
\label{eq:omega_wave}%
\end{equation}
From Eqs.~(\ref{eq:detuning}),~(\ref{eq:Oil}),~and~(\ref{eq:omega_wave}), the
detuning of the term with $N^{\prime}$, $k_{1}^{\prime}$, $k_{2}^{\prime}$ is
then given by
\begin{equation}
\Delta_{N^{\prime}}^{k_{1}^{\prime}k_{2}^{\prime}}=\left(  N^{\prime
}+1\right)  \omega_{z}+\sum_{l=1}^{2}\left(  N^{\prime}k_{l}+k_{l}^{\prime
}\right)  l_{l}\omega_{\gcd}. \label{eq:DNk1k2p}%
\end{equation}
Thus the terms with the Bessel function $J_{1}\left(  x\right)  $ will have
the detuning
\begin{equation}
\Delta_{1}^{k_{1}^{\prime}k_{2}^{\prime}}=2\omega_{z}+\sum_{l=1}^{2}\left(
k_{l}+k_{l}^{\prime}\right)  l_{l}\omega_{\gcd}.
\end{equation}
We expect that the terms with $J_{1}\left(  x\right)  $ are nonresonant. Thus,
the relation that $\Delta_{1}^{k_{1}^{\prime}k_{2}^{\prime}}\neq0$ must hold.
A simple but sufficient condition is
\begin{equation}
2\,\omega_{z}\neq k\,\omega_{\mathrm{\gcd}}, \label{eq:con1}%
\end{equation}
where $k$ is an integer. Similarly, for the terms with $J_{0}\left(  x\right)
$, $J_{2}\left(  x\right)  $, and $J_{\bar{2}}\left(  x\right)  $, the
sufficient conditions can be given by
\begin{align}
\omega_{z}  &  \neq k\,\omega_{\gcd},\label{eq:con0}\\
3\,\omega_{z}  &  \neq k\,\omega_{\mathrm{\gcd}},\label{eq:con2}\\
\omega_{z}  &  \neq k\,\omega_{\gcd}. \label{eq:con_2}%
\end{align}
The conditions in Eqs.~(\ref{eq:con1})-(\ref{eq:con_2}) can be summarized as
\begin{equation}
6\,\omega_{z}\neq k\,\omega_{\gcd}. \label{eq:Con_omegaz}%
\end{equation}
We can also assume that the longitudinal frequency $\omega_{z}$ of the qubit
is
\begin{equation}
\omega_{z}=\left(  p+r\right)  \omega_{\gcd}, \label{eq:OzForm}%
\end{equation}
where $p$ is the integer part, and $r\ $is the fraction part. To meet
Eq.~(\ref{eq:Con_omegaz}), there should be
\begin{equation}
r\notin\left\{  0,\frac{1}{6},\frac{2}{6},\frac{3}{6},\frac{4}{6},\frac{5}%
{6}\right\}  . \label{eq:con_r}%
\end{equation}

The nonresonant terms with $J_{1}$, $J_{0}$, and $J_{\pm2}$ still have effect
on the desired time evolution. These effects can be further eliminated by
decreasing the stark shifts caused by the terms with $\Omega_{N^{\prime}%
n_{1}^{\prime}n_{2}^{\prime}}^{k_{1}^{\prime}k_{2}^{\prime}}$ in
Eq.~(\ref{eq:Rabi}). The ideal case is%
\begin{equation}
\frac{\left\vert \Omega_{N^{\prime}n_{1}^{\prime}n_{2}^{\prime}}%
^{k_{1}^{\prime}k_{2}^{\prime}}\right\vert ^{2}}{|\Delta_{N^{\prime}}%
^{k_{1}^{\prime}k_{2}^{\prime}}|}\ll\left\vert \Omega_{\bar{1}n_{1}n_{2}%
}^{k_{1}k_{2}}\right\vert \text{,} \label{eq:con_Rabi0}%
\end{equation}
or equivalently,%
\begin{equation}
\frac{\left\vert \Omega_{N^{\prime}n_{1}^{\prime}n_{2}^{\prime}}%
^{k_{1}^{\prime}k_{2}^{\prime}}\right\vert ^{2}}{\left\vert \Omega_{\bar
{1}n_{1}n_{2}}^{k_{1}k_{2}}\right\vert }\ll|\Delta_{N^{\prime}}^{k_{1}%
^{\prime}k_{2}^{\prime}}|\text{,} \label{eq:con_Rabi}%
\end{equation}
for $N^{\prime}=0$, $1$, $\pm2$, $n_{1}^{\prime}+n_{2}^{\prime}\leq N_{\max}$,
and $n_{1}+n_{2}\leq N_{\max}$, where the constraint condition for $n_{l}$ and
$n_{l}^{\prime}$ denotes the working space of our algorithm.
Equation~(\ref{eq:con_Rabi0}) means that the stark shifts should be negligibly
smaller than the Rabi frequencies for state generation. Considering that
$N_{\max}$ is the maximum photon number of the target state, and using
Eq.~(\ref{eq:DNk1k2p}) and Eq.~(\ref{eq:OzForm}), we can obtain\textbf{ }%
\begin{align}
\left\vert \Delta_{0}^{k_{1}^{\prime}k_{2}^{\prime}}\right\vert  &  \geq
r\omega_{\gcd}\text{ or }\left(  1-r\right)  \omega_{\gcd}\\
\left\vert \Delta_{1}^{k_{1}^{\prime}k_{2}^{\prime}}\right\vert  &
\geq\left(  2r-\left\lfloor 2r\right\rfloor \right)  \omega_{\gcd}\text{ or
}\left(  \left\lceil 2r\right\rceil -2r\right)  \omega_{\gcd},\\
\left\vert \Delta_{2}^{k_{1}^{\prime}k_{2}^{\prime}}\right\vert  &
\geq\left(  3r-\left\lfloor 3r\right\rfloor \right)  \omega_{\gcd}\text{ or
}\left(  \left\lceil 3r\right\rceil -3r\right)  \omega_{\gcd},\\
\left\vert \Delta_{\bar{2}}^{k_{1}^{\prime}k_{2}^{\prime}}\right\vert  &  \geq
r\omega_{\gcd}\text{ or }\left(  1-r\right)  \omega_{\gcd}.
\end{align}
Here, $\left\lfloor x\right\rfloor $ means $x$ rounded down and $\left\lceil
x\right\rceil $ means $x$ rounded up. We thus reduce Eq.~(\ref{eq:con_Rabi})
to
\begin{align}
\frac{\left\vert \Omega_{0n_{1}^{\prime}n_{2}^{\prime}}^{k_{1}^{\prime}%
k_{2}^{\prime}}\right\vert ^{2}}{\left\vert \Omega_{\bar{1}n_{1}n_{2}}%
^{k_{1}k_{2}}\right\vert }  &  \ll\min\{r,1-r\}\omega_{\gcd}%
,\label{eq:con_Rabi2-1}\\
\frac{\left\vert \Omega_{1n_{1}^{\prime}n_{2}^{\prime}}^{k_{1}^{\prime}%
k_{2}^{\prime}}\right\vert ^{2}}{\left\vert \Omega_{\bar{1}n_{1}n_{2}}%
^{k_{1}k_{2}}\right\vert }  &  \ll\min\{2r-\left\lfloor 2r\right\rfloor
,\left\lceil 2r\right\rceil -2r\}\omega_{\gcd},\label{eq:con_Rabi2-2}\\
\frac{\left\vert \Omega_{2n_{1}^{\prime}n_{2}^{\prime}}^{k_{1}^{\prime}%
k_{2}^{\prime}}\right\vert ^{2}}{\left\vert \Omega_{\bar{1}n_{1}n_{2}}%
^{k_{1}k_{2}}\right\vert }  &  \ll\min\{3r-\left\lfloor 3r\right\rfloor
,\left\lceil 3r\right\rceil -3r\}\omega_{\gcd},\label{eq:con_Rabi2-3}\\
\frac{\left\vert \Omega_{\bar{2}n_{1}^{\prime}n_{2}^{\prime}}^{k_{1}^{\prime
}k_{2}^{\prime}}\right\vert ^{2}}{\left\vert \Omega_{\bar{1}n_{1}n_{2}}%
^{k_{1}k_{2}}\right\vert }  &  \ll\min\{r,1-r\}\omega_{\gcd},
\label{eq:con_Rabi2-4}%
\end{align}
a condition much stronger than Eq.~(\ref{eq:con_Rabi}). If
Eqs.~(\ref{eq:con_Rabi2-1})-(\ref{eq:con_Rabi2-4}) are fulfilled, the
nonresonant terms can in principle be suppressed. We know from
Eq.~(\ref{eq:Rabi}) that Eqs.~(\ref{eq:con_Rabi2-1})-(\ref{eq:con_Rabi2-4})
can be satisfied if, for example, the parameter $\omega_{x}$ of the qubit is
tuned sufficiently small, assuming that the reduced driving frequency $x$ and
Lamb-Dicke parameters $\eta_{l}$ have been appropriately chosen.

Beside the terms with Bessel functions $J_{N^{\prime}}(x)$ where $N^{\prime
}\neq-1$, there are also unwanted terms with the Bessel function $J_{\bar{1}%
}\left(  x\right)  $, which, however, also satisfy the resonant condition
\begin{equation}
\Delta_{\bar{1}}^{k_{1}^{\prime}k_{2}^{\prime}}=\sum_{l=1}^{2}\left(
k_{l}^{\prime}-k_{l}\right)  l_{l}\omega_{\gcd}=0. \label{eq:UnResTer}%
\end{equation}
Here, we have used Eqs.~(\ref{eq:omega_wave}) and (\ref{eq:DNk1k2p}) to obtain
Eq.~(\ref{eq:UnResTer}). The Lamb-Dicke parameters satisfy the condition
$\eta_{l}=2g_{l}/\omega_{l}\lesssim1$ for circuit QED systems even in the
ultrastrong regime~\cite{Niemczyk2010,Forn2010,Stassi2013}. From
Eq.~(\ref{eq:Oil}), we know that $l_{1}$ and $l_{2}$ are coprime numbers. We
can further make $l_{1}$ (or $l_{2}$) sufficiently large. Thus the unwanted
resonant terms will possess large $|k_{1}^{\prime}|$ (or $|k_{2}^{\prime}|$).
In this way, the effects of these terms will be suppressed due to the
exponential decrease via the term $\eta_{l}^{\left\vert k_{l}\right\vert }$ in
Eq.~(\ref{eq:Rabi}). The condition, that the term $J_{\bar{1}}\left(
x\right)  $ is negligibly small, can be summarized as that $l_{1}$ and $l_{2}$
should satisfy
\begin{equation}
\eta_{1}^{l_{2}}\eta_{2}^{l_{1}}\ll1. \label{eq:con_li}%
\end{equation}

We now summarize the condition that minimizes the effects of unwanted terms.
The parameter $\omega_{z}$ of the qubit should satisfy Eq.~(\ref{eq:OzForm})
and Eq.~(\ref{eq:con_r}). However, the parameter $\omega_{x}$ of the qubit is
mainly constrained by current experiments. For example, typical values of
$\omega_{x}/2\pi$ are in the range $1\sim5$ GHz. The frequencies of the cavity
modes $\omega_{l}$ should satisfy Eq.~(\ref{eq:Oil}) and Eq.~(\ref{eq:con_li}%
). The values of the reduced driving frequency $x=2\Omega/\tilde{\omega}$ and
Lamb-Dick parameter $\eta_{l}=2g_{l}/\omega_{l}$ should satisfy
Eq.~(\ref{eq:con_Rabi}) or stronger conditions Eqs.~(\ref{eq:con_Rabi2-1}%
)-(\ref{eq:con_Rabi2-4}). Appropriate values of $x$ and $\eta_{l}$ can be
obtained via numerical simulations, which will be discussed below in
Sec.~\ref{sec:UnTer}.

\subsection{Numerical simulations\label{sec:UnTer}}

We now further numerically simulate the effect of the unwanted terms on the
generation of target states by using examples of generating the following two
target states
\begin{align}
\left\vert \tilde{\psi}_{\mathrm{E}}\right\rangle  &  =\sum_{n_{1}+n_{2}\leq
2}\frac{1}{\sqrt{6}}\left\vert n_{1},n_{2}\right\rangle \left\vert
g\right\rangle ,\label{eq:psiG}\\
\left\vert \tilde{\psi}_{\text{\textrm{N}}}\right\rangle  &  =\frac{1}%
{\sqrt{2}}\left(  \left\vert 0,2\right\rangle +\left\vert 2,0\right\rangle
\right)  \left\vert g\right\rangle , \label{eq:psiNOON}%
\end{align}
for some given parameters. It is obvious that $|\tilde{\psi}_{\text{E}}%
\rangle$ is an entangled state where every state component is evenly occupied.
We thus call $|\tilde{\psi}_{\text{E}}\rangle$ the evenly-populated state. The
state $|\tilde{\psi}_{\text{N}}\rangle$ is a two-photon NOON
state~\cite{Nielsen2010}. Both $|\tilde{\psi}_{\text{E}}\rangle$ and
$|\tilde{\psi}_{\text{N}}\rangle$ possess a maximum photon number $N_{\max}%
=2$. The fidelities for generating these two states $|\tilde{\psi}_{\text{E}%
}\rangle$ and $|\tilde{\psi}_{\text{N}}\rangle$ are defined as
\begin{align}
\mathcal{F}_{\mathrm{E}}  &  =\left\vert \left\langle \tilde{\psi}%
_{\mathrm{E}}^{A}\right.  \left\vert \tilde{\psi}_{\mathrm{E}}\right\rangle
\right\vert ,\\
\mathcal{F}_{\mathrm{N}}  &  =\left\vert \left\langle \tilde{\psi}%
_{\mathrm{N}}^{A}\right.  \left\vert \tilde{\psi}_{\mathrm{N}}\right\rangle
\right\vert .
\end{align}
Here, $|\tilde{\psi}_{\mathrm{E}}^{A}\rangle$ and $|\tilde{\psi}_{\mathrm{N}%
}^{A}\rangle$ are respectively the actually generated states via the total
Hamiltonian in Eq.~(\ref{eq:H}).

We now determine the detailed experimental parameters. From
Eq.~(\ref{eq:OzForm}) and Eq.~(\ref{eq:con_r}), we set $r=$ $3/4$, $p=9$, and
$\omega_{\gcd}/2\pi=2$ GHz, which corresponds to $\omega_{z}/2=19.5$ GHz. From
Eq.~(\ref{eq:con_Rabi}) or Eqs.~(\ref{eq:con_Rabi2-1})-(\ref{eq:con_Rabi2-4}),
the parameter $\omega_{x}/2\pi$ should be made smaller, e.g., we set
$\omega_{x}/2\pi=1.2$ GHz. The frequency of the $l$th cavity, i.e.,
$\omega_{l}$, is determined by Eq.~(\ref{eq:Oil}) and (\ref{eq:con_li}). Since
the microwave fields are usually of several gigahertz, here we set $l_{1}=3$,
and $l_{2}=4$, thus yielding $\omega_{1}/2\pi=l_{1}\omega_{\gcd}/2\pi=6$ GHz
and $\omega_{2}/2\pi=l_{2}\omega_{\gcd}/2\pi=8$ GHz. The Lamb-Dicke parameters
for the first and second cavity modes are set to be identical, i.e.,
\begin{equation}
\eta_{1}=\eta_{2}=\eta.
\end{equation}

We vary the Lamb-Dick parameter $\eta$ and the reduced driving frequency
$x=2\Omega/\tilde{\omega}$ to simulate the effect of the unwanted terms on the
fidelity of the expected target states in Eqs.~(\ref{eq:psiG}) and
(\ref{eq:psiNOON}). The pulses are taken according to the
calculation of Chapter III B. That is, in the $\nu$th step, we use a
sinusoidal driving with the driving frequency $\tilde{\omega}_{\nu}$. Since
the sinusoidal driving lasts for a duration $t_{\nu}$, the driving field can
be considered as square-windowed sinusoidal signal and thus, strictly
speaking, is not delta-shaped in the spectrum. The simulation
results for generating target states in Eq.~(\ref{eq:psiG}) and
Eq.~(\ref{eq:psiNOON}) are listed in Table~\ref{tab:psiG} and
Table~\ref{tab:psiNOON}, respectively. We can easily find that larger reduced
driving strengths $x$ and Lamb-Dick parameters $\eta$ can usually make the
fidelity higher. For the evenly-populated state $|\tilde{\psi}_{\mathrm{E}%
}\rangle$, the largest fidelity 0.939 can be obtained at $x=1.7571$ and
$\eta=0.3714$. However, for the NOON state $|\tilde{\psi}_{\mathrm{N}}\rangle
$, the largest fidelity 0.92 can be obtained at $x=2$ and $\eta=0.5429$.%

\begin{table}[tbp] \centering
\caption{The fidelities $\mathcal{F}_{\mathrm{E}}=\vert\langle \tilde{\psi}_{\mathrm{E}}^{A}
\vert \tilde{\psi}_{\mathrm{E}}\rangle\vert$ of the target state
$\vert \tilde{\psi}_{\mathrm{E}}\rangle
=\sum_{n_1+n_2\le 2}
(1/\sqrt{6})\vert n_{2},n_{2}\rangle\vert g\rangle$ are listed for different values of
the reduced driving frequency $x=2\Omega/\tilde{\omega}$ and the Lamb-Dicke
parameter $\eta=2g_1/\omega_1=2g_2/\omega_2$. Here $\vert\tilde{\psi}_{\mathrm{E}}^{A}\rangle$ is
the actually generated state using the total Hamiltonian. We have chosen
the longitudinal frequency of the qubit $\omega_{z}/2\pi=19.5\operatorname{GHz}$,
the transverse frequency of the qubit $\omega_{x}/2\pi=1.2\operatorname{GHz}$,
the frequency of the first mode $\omega_1/2\pi=6\operatorname{GHz}$ and the frequency of the second mode
$\omega_2/2\pi=8\operatorname{GHz}$.}\qquad%

\begin{tabular}
[c]{c|c|cccccc}\hline\hline
\multicolumn{2}{c|}{} & \multicolumn{6}{|c}{$\eta$}\\\cline{3-8}%
\multicolumn{2}{c|}{} & 0.2 & 0.3714 & 0.4571 & 0.5429 & 0.6286 &
0.7143\\\hline
& 0.3 & 0.115 & 0.393 & 0.519 & 0.608 & 0.739 & 0.675\\
& 0.7857 & 0.589 & 0.817 & 0.872 & 0.851 & 0.911 & 0.852\\
$x$ & 1.0286 & 0.615 & 0.85 & 0.876 & 0.906 & 0.886 & 0.857\\
& 1.2714 & 0.724 & 0.872 & 0.886 & 0.906 & 0.852 & 0.878\\
& 1.7571 & 0.821 & 0.939 & 0.899 & 0.859 & 0.825 & 0.837\\
& 2 & 0.867 & 0.915 & 0.838 & 0.876 & 0.887 & 0.859\\\hline\hline
\end{tabular}
\label{tab:psiG}%
\end{table}%
%

\begin{table}[tbp] \centering
\caption{The fidelities $\mathcal{F}_{\mathrm{N}}=\vert\langle \tilde{\psi
}_{\mathrm{N}}^{A}\vert \tilde{\psi}_{\mathrm{N}}\rangle \vert$ of the target state $\vert \tilde{\psi}_{\mathrm{NOON}}\rangle=
(1/\sqrt{2})(\vert 0,2\rangle\vert g\rangle+\vert 2,0\rangle\vert g\rangle)$ are listed for different values of
the reduced driving frequency $x=2\Omega/\tilde{\omega}$ and the Lamb-Dicke
parameter $\eta=2g_1/\omega_1=2g_2/\omega_2$. Here $\vert\tilde{\psi}_{\mathrm{G}}^{A}\rangle$ is
the actually generated state using the total Hamiltonian. We have chosen
the same parameters as in Table.~\ref{tab:psiG}.}\qquad%

\begin{tabular}
[c]{c|c|cccccc}\hline\hline
\multicolumn{2}{c|}{} & \multicolumn{6}{|c}{$\eta$}\\\cline{3-8}%
\multicolumn{2}{c|}{} & 0.2 & 0.3714 & 0.4571 & 0.5429 & 0.6286 &
0.7143\\\hline
& 0.3 & 0.108 & 0.34 & 0.403 & 0.395 & 0.461 & 0.687\\
& 0.7857 & 0.675 & 0.815 & 0.833 & 0.862 & 0.829 & 0.868\\
$x$ & 1.0286 & 0.78 & 0.857 & 0.846 & 0.867 & 0.883 & 0.813\\
& 1.5143 & 0.871 & 0.877 & 0.873 & 0.877 & 0.787 & 0.806\\
& 1.7571 & 0.844 & 0.918 & 0.889 & 0.902 & 0.862 & 0.819\\
& 2 & 0.876 & 0.909 & 0.832 & 0.92 & 0.853 & 0.806\\\hline\hline
\end{tabular}
\label{tab:psiNOON}%
\end{table}%

\section{Environmental effect on target states \label{sec:SimResults}}

In\textbf{ }the above, we only discuss the effect of unwanted terms on the
generation of target states. We now study the effect of dissipation on the
fidelities of target states by numerical simulation for given parameters. When
the environmental effect is included, the dynamical evolution of the SQC can
be described by the master equation
\begin{align}
\rho=  &  -i\left[  H,\rho\right]  +\mathcal{D}\left[  \sqrt{\gamma_{eg}%
}\tilde{\sigma}_{ge}\right]  \rho+\mathcal{D}\left[  \sqrt{\gamma_{ee}}%
\tilde{\sigma}_{ee}\right]  \rho\nonumber\\
&  +\mathcal{D}\left[  \sqrt{\gamma_{gg}}\tilde{\sigma}_{gg}\right]
\rho+\mathcal{D}\left[  \sqrt{\kappa_{1}}a_{1}\right]  \rho+\mathcal{D}\left[
\sqrt{\kappa_{2}}a_{2}\right]  \rho, \label{eq:ME_Schr}%
\end{align}
where $\rho$ and $H$ are the reduced density operator and the Hamiltonian of
the whole system, respectively. The total Hamiltonian has been given in
Eq.~(\ref{eq:H}). The compact notation $\mathcal{D}\left[  c\right]
\rho=\left(  2c\rho c^{\dag}-c^{\dag}c\rho-\rho c^{\dag}c\right)  /2$
represents the Lindblad-type dissipation. We have noted that $\{|g\rangle
,|e\rangle\}$ is the basis of $\sigma_{z}$, but the qubit dissipation is
determined by the qubit basis $\left\{  |\tilde{g}\rangle,|\tilde{e}%
\rangle\right\}  $. The ground ($|\tilde{g}\rangle$) and excited ($|\tilde
{e}\rangle$) states of the qubit are given by the eigenstates of
Eq.~(\ref{eq:Hq}). If we define
\begin{equation}
\sigma_{\nu\mu}=\left\vert \nu\right\rangle \left\langle \mu\right\vert ,
\end{equation}
with $\nu=g,e$ and $\mu=g,e$, and also define
\begin{equation}
\tilde{\sigma}_{\nu\mu}=\left\vert \tilde{\nu}\right\rangle \left\langle
\tilde{\mu}\right\vert ,
\end{equation}
with $\tilde{\nu}=\tilde{g},\tilde{e}$ and $\tilde{\mu}=\tilde{g},\tilde{e}$.
We can easily verify
\[
\tilde{\sigma}_{\nu\mu}=R_{y}\left(  \theta\right)  \sigma_{\nu\mu}R_{y}%
^{\dag}\left(  \theta\right)  ,
\]
where $R_{y}\left(  \theta\right)  =\exp\left(  -i\theta\sigma_{y}/2\right)
$, and $\theta=\arctan(\omega_{x}/\omega_{z})$. In Eq.~(\ref{eq:ME_Schr}),
$\gamma_{eg}$ is the pure-relaxation rate from the qubit excited state to the
ground state. Besides, $\gamma_{gg}$ and $\gamma_{ee}$ are the pure-dephasing
rates originating from disturbed qubit eigenstates. The decay rates of the
first and the second cavity fields are denoted by $\kappa_{1}$ and $\kappa
_{2}$, respectively.

Using parameters in Sec.~\ref{sec:UnTer} and taking the reduced
driving\textbf{ }strength $x=2\Omega/\tilde{\omega}=1.7571$ and Lamb-Dicke
parameter $\eta=0.3714$ from Table~\ref{tab:psiG} and Table~\ref{tab:psiNOON},
we find that the highest fidelity $\mathcal{F}_{\mathrm{E}}=0.939$ is achieved
for generating the evenly-populated state $|\tilde{\psi}_{\mathrm{E}}\rangle$
in Eq.~(\ref{eq:psiG}), and a high fidelity $\mathcal{F}_{\mathrm{N}}=0.918$
is also reached for generating the NOON state $|\tilde{\psi}_{\mathrm{N}%
}\rangle$ in Eq.~(\ref{eq:psiNOON}).

We now assume that the decay rates in Eq.~(\ref{eq:ME_Schr}) are taken as
$\gamma_{gg}/2\pi=0$, $\gamma_{ee}/2\pi=2$ MHz, and $\gamma_{eg}/2\pi
=\kappa_{1}/2\pi=\kappa_{2}/2\pi=1$ MHz. We assume that the density operators
$\rho_{\mathrm{E}}^{A}$ and $\rho_{\mathrm{N}}^{A}$ are the actually generated
states for the target states $|\tilde{\psi}_{\mathrm{E}}\rangle$ and
$|\tilde{\psi}_{\mathrm{N}}\rangle$. Then the fidelities can be redefined as
\begin{align}
\mathcal{F}_{\mathrm{E}}^{\prime}  &  =\sqrt{\left\langle \tilde{\psi
}_{\mathrm{E}}\right\vert \rho_{\mathrm{E}}^{A}\left\vert \tilde{\psi
}_{\mathrm{E}}\right\rangle },\\
\mathcal{F}_{\mathrm{N}}^{\prime}  &  =\sqrt{\left\langle \tilde{\psi
}_{\mathrm{N}}\right\vert \rho_{\mathrm{N}}^{A}\left\vert \tilde{\psi
}_{\mathrm{N}}\right\rangle }.
\end{align}
We perform numerical simulations using the above parameters and obtain
$\mathcal{F}_{\mathrm{E}}^{\prime}=$ $0.911$ and $\mathcal{F}_{\mathrm{N}%
}^{\prime}=0.863$. The total time for generating $|\tilde{\psi}_{\mathrm{E}%
}\rangle$ is $T_{\mathrm{E}}=8.9561%
\operatorname{ns}%
$ and that for generating $|\tilde{\psi}_{\mathrm{N}}\rangle$ is
$T_{\mathrm{N}}=10.4451%
\operatorname{ns}%
$. Both $T_{E}$ and $T_{\mathrm{N}}$ are too small to induce significant
decoherence at the decay rates specified by us. Thus, the fidelity losses
induced by dissipation are fairly small, which are $\mathcal{F}_{\mathrm{E}%
}-\mathcal{F}_{\mathrm{E}}^{\prime}=0.028$ for the evenly-populated state
$|\tilde{\psi}_{E}\rangle$ and $\mathcal{F}_{\mathrm{N}}-\mathcal{F}%
_{\mathrm{N}}^{\prime}=0.055$ for the NOON state $|\tilde{\psi}_{\mathrm{N}%
}\rangle$.

\section{Discussions\label{sec:discussions}}

We now discuss the advantages and disadvantages between our methods and the
previous
ones~\cite{Gardiner1997,Drobny1998,Zheng2000,Kneer1998,Strauch2010,Strauch2012PRA,Sharma2015,Zou2002}
for generating arbitrarily entangled states of two microwave fields or two
vibrational modes.
\begin{table}[tbp] \centering%
\caption{Comparison of different methods for generating arbitrarily entangled states of two-mode bosonic fields.
We use Pop. Leak., St. Eff. and Mult. Proc. to denote population leakage, the Stark effect, multiboson processes, respectively. No. At. Lev. is used to denote the number of atomic energy levels. For example, $2$ denotes two energy levels
when the state is generated. We use "Yes" or "No" to denote whether the population leakage (Stark effect) occurs (are used) or not. Meanwhile, L (or H). Pn. No. means multiphonon processes of low (or high) phonon number,
however, L (or H). Pt. No. means multiphoton processes of low (or high) photon number.}
\begin{tabular}
[c]{c|c|c|c|c}\hline\hline
& \textbf{Pop. Leak.} & \textbf{No. At. Lev.} & \textbf{St. Eff.} &
\textbf{Mult. Proc. }\\\hline
Ref.~\cite{Gardiner1997} & Yes & 2 & No & L. Pn. No.\\
Ref.~\cite{Drobny1998} & No & 3 & No & L. Pn. No.\\
Ref.~\cite{Zheng2000} & No & 3 & No & H. Pn. No.\\
Ref.~\cite{Kneer1998} & No & 2 & Yes & None\\
Ref.~\cite{Zou2002} & No & 2 & No & H. Pn. No.\\
Refs.~\cite{Strauch2010,Strauch2012PRA,Sharma2015} & No & 2 & Yes & None\\
Our proposal & No & 2 & No & L. Pt. No.\\\hline\hline
\end{tabular}
\label{tab:CompareUniversal}%
\end{table}%

The brief comparison between these methods is listed in
Table.~\ref{tab:CompareUniversal}. In detail, Ref.~\cite{Gardiner1997}
provided an algorithm to generate arbitrarily entangled states of two
vibrational modes. But due to population leakage outside the original space, it
takes an exponential complexity of the number of steps. The succeeding
proposals~\cite{Drobny1998,Zheng2000,Kneer1998,Strauch2010,Strauch2012PRA,Sharma2015,Zou2002}
overcome the exponential drawback in several ways: (1) A third atomic level is
used to shield oscillations that cause population leakage~\cite{Drobny1998,Zheng2000}. But the
disadvantage is that higher energy levels of systems usually have larger decay
rates, which inevitably reduce the fidelities of the target states. (2)
Boson-number-dependent Stark effects are used to realize independent
operations of particular
states~\cite{Kneer1998,Strauch2010,Strauch2012PRA,Sharma2015}. But the
disadvantage is that the detunings of nonresonant terms are usually less by
one order of the coupling strengths between the two-level system and boson
modes. This means that the Rabi frequencies are smaller, and the longer
generation time is required. (3) Multiphoton processes of high photon number
are used to shield oscillations that cause population leakage or reduce the number of
steps~\cite{Zou2002,Zheng2000}. But the disadvantage is that if the coupling
strengths between the atom and cavity fields are not high enough, then the
Rabi frequencies become small, especially for states with high photon numbers,
which obviously indicates longer generation time.

Besides the advantage that there is no population leakage, our method has also
the following advantages compared with previous
ones.~\cite{Drobny1998,Zheng2000,Kneer1998,Strauch2010,Strauch2012PRA,Sharma2015,Zou2002}%
: (1) It only uses the two energy levels of the qubit. Thus, the fidelities of
the target states should be higher because there is no other auxiliary energy
levels. (2) The detunings of the nonresonant terms are in the order of the
resonator frequencies. They are usually bigger than the coupling strengths
between the qubit and resonator modes. Thus the Rabi frequency can be made
bigger than those using boson-number-dependent Stark effects. (3) We use
multiphoton processes of low photon number, i.e., one photon at most is
converted for either mode. Thus the Rabi frequency can be bigger than those
using multiphoton processes of higher photon number,especially when the
coupling strengths between the qubit and cavity modes are not very big. Of
course, stronger couplings will further enhance the Rabi frequencies and hence
reduce the generation time.

We point out that the real supercoducting qubit circuits are mutilevel
systems, the information leakage to higher levels is not avoidable. However,
the leakage can be neglected when the transition frequency between the first
excited state and the second excited state is much larger than the qubit
frequency. For example, in the flux qubit circuits, due to its large
anharmonicity of energy levels, the information leakage is negligibly small.
However, for the transmon and phase qubit, the anharmonicity is very weak.
Thus, the pulse should be carefully calibrated to avoid information leakage to
higher levels. The pulse calibration can be done as in Ref.~\cite{Steffen2003}%
.

We now compare the differences between our algorithm and other ones for
generating NOON states. Ref.~\cite{Zou2001} uses mutliphoton processes to
generate NOON states. In superconducting systems, this means a low generation
efficiency if the Lamb-Dicke parameter is not sufficiently big.
Ref.~\cite{Xu2013} uses synchronization technology to generate NOON states,
but the time duration for synchronization between two steps can be quite long
and there exists inevitably information leakage. Ref.~\cite{Merkel2010} and
its experimental realization~\cite{H.Wang2011} use two phase qubits with three
active energy levels to generate NOON states of two cavity modes. The
experimental setup is complex and the high energy levels of qubits will reduce
the decoherence time.\textbf{ }Ref.~\cite{Strauch2010} uses photon
number-dependent Stark effects to achieve independent operations. Thus the
Rabi frequency is smaller than the qubit-cavity coupling strengths.
Ref.~\cite{Strauch2012} requires that two qubits be initially prepared in a
Bell state and finally get decoupled from the qubits and cavity fields.
Ref.~\cite{Su2014} uses one qubit but still needs one additional level to
shield unwanted resonances. More recently, Ref.~\cite{Xiong2015} uses one
qubit of four levels which resonantly interacts with two resonators
simultaneously to speed up the generation process of NOON states.

When applied to generating NOON states, our algorithm has new features besides
the common advantages for generating arbitrary two-mode photon states: (1)
Only carrier processes~\cite{trappedions} and one-photon processes are used.
In this case, even though the coupling strengths between the qubit and cavity
modes are small, large Rabi frequencies can still be obtained. (2) The number
of steps is reduced to linear dependence on the maximum photon number. These
advantages indicate less generation time and thus guarantee a higher
efficiency than preceding methods.

Now we discuss the experimental feasibility of our scheme.
Table~\ref{tab:psiG} and Table~\ref{tab:psiNOON} show that without pulse
calibration, higher fidelities can be achieved at bigger\textbf{ }Lamb-Dicke
parameters $\eta$ and reduced driving frequencies $x$. These values are
already in the ultrastrong regime. Ref.~\cite{Niemczyk2010} has reported
ultrastrong couplings between three resonator modes and a flux qubit, where
the Lamb-Dicke parameter $\eta$ can reach as high as $0.236$. In the
ultrastrong regime, Rabi frequencies can be made to approach the magnitude of
$\omega_{x}$, which usually ranges from $1$ to $5$ GHz. The decay rates of the
qubit and cavity fields are usually in the magnitude of megahertz. Thus the
dissipation has small effect on the fidelities of target states. For
singe-mode microwave fields, Fock states with up to six
photons~\cite{Martinis-1} and Fock state superpositions~\cite{Martinis-2} have
been experimentally demonstrated using phase qubits. The NOON state up to $3$
photons has also been experimentally reported ~\cite{H.Wang2011}. We thus hope
that our proposal is also experimentally feasible in the near future.

\section{Conclusions\label{sec:conclusions}}

In summary, we have proposed an approach to generate arbitrary superpositions
of photon states of two microwave fields in two separated cavities. Our method
mainly depends on the coexistence of transverse and longitudinal couplings
between the qubit and cavity fields. Employing the longitudinal couplings, we
derive a Hamiltonian which is similar to that of trapped ions interacting with
two vibrational modes~\cite{Steinbach1997}. Using four simple interaction
Hamiltonians derived from the longitudinal coupling, we design the state
generation algorithm. Our algorithm can be regarded as the improved version of
that in~\cite{Gardiner1997} when the transverse and longitudinal couplings
coexist in circuit QED systems. But it has remedied the drawback that the
number of steps exponentially depends on the maximal photon number, which is
replaced by a quadratic dependence. Compared with previous ones with quadratic
complexity, our algorithm does not require atomic energy levels higher than
two~\cite{Drobny1998,Zheng2000}, boson-number-dependent stark
effects~\cite{Kneer1998,Strauch2010,Sharma2015}, or multiboson processes of
high boson numbers~\cite{Zou2002,Zheng2000}.

When applied to the generation of NOON states, whose engineering has been
extensively
studied~\cite{Zou2001,Xu2013,CPYang2003,Merkel2010,H.Wang2011,Strauch2012,Su2014,Xiong2015,Peng2012}%
, our algorithm needs only carrier and one-photon sideband transitions.
Meanwhile, the number of steps only linearly depends on the maximum photon
numbers. In fact, these properties for generating NOON states can be
generalized to any states with a constant total photon number of both modes.

We have also discussed how to avoid the effect of unwanted terms on the
generation of target state. Our numerical results show that fidelities above
$0.91$ can be reached in the ultrastrong regime for the two-photon
evenly-populated state and NOON state when the environmental effect is
neglected. The generation time can be very short, in which case, the
environment has small effect on fidelities of the target states. We here note
that due to the similarity of two-mode interaction Hamiltonians, the algorithm
using two-mode multi-phonon processes in Ref.~\cite{Zou2002} can be directly
applied into our model. Thus, two-mode Fock states with high photon numbers
can be generated with just two steps as one-mode Fock states
in~\cite{Zhao2015}.

We have noted that our method for generating NOON states is similar to a
recent algorithm simplified from the one which employs Stark effects to
generate arbitrary entangled states~\cite{Sharma2015}.

\section{Acknowledgement}

YXL is supported by the National Basic Research Program of China Grant
No.~2014CB921401, the NSFC Grants No.~61025022, and No.~91321208.

\appendix

\section{Universal algorithm for arbitrary maximum photon
numbers\label{append:Univ_alg}}

The basic principle of our algorithm is state space reduction. The state space
of the target state in Eq.~(\ref{eq:psi_f}) can be denoted\ as
\begin{equation}
\mathcal{H}_{2N_{\max}}=\left\{  \left\vert n_{1},n_{2}\right\rangle
\left\vert g\right\rangle |n_{1}+n_{2}\leq N_{\max}\right\}  .
\end{equation}
We will implement $2N_{\max}$ procedures, with each procedure containing some steps.

In the 1st procedure, we aim to clear the populations in the subspace
\begin{equation}
\mathcal{H}_{2N_{\max}}^{\prime}=\left\{  \left\vert n_{1},n_{2}\right\rangle
\left\vert g\right\rangle |n_{1}+n_{2}=N_{\max}\right\}
\end{equation}
of $\mathcal{H}_{2N_{\max}}$. This can be achieved via alternatively switching
$N_{\max}$ \textquotedblleft\={1}0\textquotedblright\ transitions and
$N_{\max}$ \textquotedblleft0\={1}\textquotedblright\ transitions,
i.e.,
\begin{align}
&  \left\vert 0,N_{\max}\right\rangle \left\vert g\right\rangle
\xrightarrow[f]{0\bar{1}}\left\vert 0,N_{\max}-1\right\rangle \left\vert
e\right\rangle \xrightarrow[f-1]{\bar{1}0}\left\vert 1,N_{\max}-1\right\rangle
\left\vert g\right\rangle \cdots\nonumber\\
&  \xrightarrow[f-2N_{\mathrm{max}}+2]{0\bar{1}}\left\vert N_{\max
}-1,0\right\rangle \left\vert e\right\rangle
\xleftarrow[f-2N_{\mathrm{max}}+1]{\bar{1}0}\left\vert N_{\max},0\right\rangle
\left\vert g\right\rangle .
\end{align}
to transfer the populations in the subspace $\mathcal{H}%
_{2N_{\max}}^{\prime}$ of $\mathcal{H}_{2N_{\max}}$ to the state $\left\vert
N_{\max}-1,0\right\rangle \left\vert e\right\rangle $. Thus, $\mathcal{H}%
_{2N_{\max}}$ is reduced to the state space $\mathcal{H}_{2N_{\max}-1}$ where%
\begin{align}
\mathcal{H}_{2N_{\max}-1}=  &  \left\{  \left\vert n_{1},n_{2}\right\rangle
\left\vert g\right\rangle |n_{1}+n_{2}\leq N_{\max}-1\right\} \nonumber\\
&  \cup\left\{  \left\vert n_{1},n_{2}\right\rangle \left\vert e\right\rangle
|n_{1}+n_{2}\leq N_{\max}-2\right\} \nonumber\\
&  \cup\left\{  \left\vert N_{\max}-1,0\right\rangle \left\vert e\right\rangle
\right\}  .
\end{align}

\begin{center}
$\vdots$
\end{center}

In the $2\mu$th procedure, we aim to clear the populations in the subspace
$\left\{  \left\vert N_{\max}-\mu,0\right\rangle \left\vert e\right\rangle
\right\}  $ of $\mathcal{H}_{2N_{\max}-2\mu+1}$. This can be achieved via
alternatively switching $N_{\max}-\mu$ \textquotedblleft1\={1}%
\textquotedblright\ transitions and $N_{\max}-\mu$ \textquotedblleft%
00\textquotedblright\ transitions, i.e.,%
\begin{align}
&  \left\vert 0,N_{\max}-\mu\right\rangle \left\vert g\right\rangle
\xrightarrow[f-N_{2\mu}]{1\bar{1}}\left\vert 1,N_{\max}-\mu-1\right\rangle
\left\vert e\right\rangle \nonumber\\
&  \xrightarrow[f-N_{2\mu}-1]{00}\left\vert 1,N_{\max}-\mu-1\right\rangle
\left\vert g\right\rangle \cdots\nonumber\\
&  \xrightarrow[f-N_{2\mu}-2(N_{\mathrm{max}}-\mu)+2]{1\bar{1}}\left\vert
N_{\max}-\mu,0\right\rangle \left\vert e\right\rangle \nonumber\\
&  \xrightarrow[f-N_{2\mu}-2(N_{\mathrm{max}}-\mu)+1]{00}\left\vert N_{\max
}-\mu,0\right\rangle \left\vert g\right\rangle .
\end{align}
with $N_{2\mu}=2N_{\max}+\left(  4N_{\max}-2\mu-1\right)  \left(
\mu-1\right)  $, to transfer the populations in the subspace
\begin{align}
\mathcal{H}_{2N_{\max}-2\mu+1}^{\prime}=  &  \left\{  \left\vert n_{1}%
,n_{2}\right\rangle \left\vert g\right\rangle |n_{1}+n_{2}=N_{\max}-\mu
,n_{2}\neq0\right\} \nonumber\\
&  \cup\left\{  \left\vert N_{\max}-\mu,0\right\rangle \left\vert
e\right\rangle \right\}
\end{align}
of $\mathcal{H}_{2N_{\max}-2\mu+1}$ to the state $\left\vert N_{\max}%
-\mu,0\right\rangle \left\vert g\right\rangle $. Here Thus, $\mathcal{H}%
_{2N_{\max}-2\mu+1}$ is reduced to the state space $\mathcal{H}_{2N_{\max
}-2\mu}$ where
\begin{align}
\mathcal{H}_{2N_{\max}-2\mu}=  &  \left\{  \left\vert n_{1},n_{2}\right\rangle
\left\vert g\right\rangle |n_{1}+n_{2}\leq N_{\max}-\mu-1\right\} \nonumber\\
&  \cup\left\{  \left\vert n_{1},n_{2}\right\rangle \left\vert e\right\rangle
|n_{1}+n_{2}\leq N_{\max}-\mu-1\right\} \nonumber\\
&  \cup\left\{  \left\vert N_{\max}-\mu,0\right\rangle \left\vert
g\right\rangle \right\}  .
\end{align}

In the $\left(  2\mu+1\right)  $th procedure, we aim to clear the populations
in the subspace $\left\{  \left\vert N_{\max}-\mu,0\right\rangle \left\vert
g\right\rangle \right\}  $ of $\mathcal{H}_{2N_{\max}-2\mu}$. This can be
achieved via alternatively switching $N_{\max}-\mu$ \textquotedblleft%
\={1}0\textquotedblright\ transitions and $N_{\max}-\mu-1$ \textquotedblleft%
0\={1}\textquotedblright\ transitions, i.e.,
\begin{align}
&  \left\vert 0,N_{\max}-\mu-1\right\rangle \left\vert e\right\rangle
\xrightarrow[f-N_{2\mu+1}]{\bar{1}0}\left\vert 1,N_{\max}-\mu-1\right\rangle
\left\vert g\right\rangle \nonumber\\
&  \xrightarrow[f-N_{2\mu+1}-1]{0\bar{1}}\left\vert 1,N_{\max}-\mu
-2\right\rangle \left\vert e\right\rangle \cdots\nonumber\\
&  \xrightarrow[f-N_{2\mu+1}-2(N_{\mathrm{max}}-\mu)+3]{\bar{1}0}\left\vert
N_{\max}-\mu-1,0\right\rangle \left\vert e\right\rangle \nonumber\\
&  \xleftarrow[f-N_{2\mu+1}-2(N_{\mathrm{max}}-\mu)+2]{0\bar{1}}\left\vert
N_{\max}-\mu,0\right\rangle \left\vert g\right\rangle .
\end{align}
with $N_{2\mu+1}=N_{2\mu}+2(N_{\max}-\mu)$, to transfer the
populations in the subspace
\begin{align}
\mathcal{H}_{2N_{\max}-2\mu}^{\prime}=  &  \left\{  \left\vert n_{1}%
,n_{2}\right\rangle \left\vert e\right\rangle |n_{1}+n_{2}=N_{\max}%
-\mu-1,n_{2}\neq0\right\} \nonumber\\
&  \cup\left\{  \left\vert N_{\max}-\mu,0\right\rangle \left\vert
g\right\rangle \right\}
\end{align}
\ of $\mathcal{H}_{2N_{\max}-2\mu}$ to the state $\left\vert N_{\max}%
-\mu-1,0\right\rangle \left\vert e\right\rangle $. Thus, $\mathcal{H}%
_{2N_{\max}-2\mu}$ is reduced to the state space $\mathcal{H}_{2N_{\max}%
-2\mu-1}$ where%
\begin{align}
\mathcal{H}_{2N_{\max}-2\mu-1}=  &  \left\{  \left\vert n_{1},n_{2}%
\right\rangle \left\vert g\right\rangle |n_{1}+n_{2}\leq N_{\max}%
-\mu-1\right\} \nonumber\\
&  \cup\left\{  \left\vert n_{1},n_{2}\right\rangle \left\vert e\right\rangle
|n_{1}+n_{2}\leq N_{\max}-\mu-2\right\} \nonumber\\
&  \cup\left\{  \left\vert N_{\max}-\mu-1,0\right\rangle \left\vert
e\right\rangle \right\}  .
\end{align}

\begin{center}
$\vdots$
\end{center}

In the ($2N_{\max}$)th procedure, we aim to clear the populations in the
subspace
\begin{equation}
\mathcal{H}_{1}^{\prime}=\left\{  \left\vert 0,0\right\rangle \left\vert
e\right\rangle \right\}
\end{equation}
of $\mathcal{H}_{1}$. This can be achieved via switching one \textquotedblleft%
00\textquotedblright\ transitions, i.e.,%
\begin{equation}
\left\vert 0,0\right\rangle \left\vert e\right\rangle
\xrightarrow[1]{00}\left\vert 0,0\right\rangle \left\vert g\right\rangle .
\end{equation}
to transfer the populations in the subspace $\mathcal{H}%
_{1}^{\prime}$ of $\mathcal{H}_{1}$ to the state $\left\vert 0,0\right\rangle
\left\vert g\right\rangle $. Thus, $\mathcal{H}_{1}$ is reduced to the state
space $\mathcal{H}_{0}$ where
\begin{equation}
\mathcal{H}_{0}=\left\{  \left\vert 0,0\right\rangle \left\vert g\right\rangle
\right\}  ,
\end{equation}
is namely the intial state space for $\left\vert \psi_{0}\right\rangle $ in
Eq.~(\ref{eq:psi_0}).

Base on the above discussion, we now calculate the number of steps to generate
the target state in Eq.~(\ref{eq:psi_f}) with an arbitrary maximum photon
number $N_{\max}$. For alternatively switching \textquotedblleft1\={1}" and
\textquotedblleft00" transitions, we need $f_{1\bar{1},00}$ steps, given by
\begin{equation}
f_{1\bar{1},00}=1+\sum_{N=1}^{N_{\max}-1}2N=N_{\max}^{2}-N_{\max}+1.
\end{equation}
For alternatively switching \textquotedblleft\={1}0" and \textquotedblleft%
0\={1}" transitions, we need $f_{\bar{1}0,0\bar{1}}$ steps, given by
\begin{equation}
f_{\bar{1}0,0\bar{1}}=1+\sum_{N=1}^{N_{\max}}\left(  2N-1\right)  =N_{\max
}^{2}+1.
\end{equation}
Therefore, the total step number $f$ is
\begin{equation}
f=f_{1\bar{1},00}+f_{\bar{1}0,0\bar{1}}=2N_{\max}^{2}-N_{\max}+2.
\label{eq:f_general}%
\end{equation}
for generating the target state in Eq.~(\ref{eq:psi_f}).

\end{document}